\documentclass[12pt]{iopart}

\usepackage{bm}
\usepackage{amssymb}  
\usepackage{centernot}
\usepackage{graphicx}
\usepackage{cite}
\usepackage[dvipsnames]{xcolor}
 
\begin{document}

\title[]{Dynamical Theory for Adaptive Systems}

\author{Tuan Minh Pham and Kunihiko Kaneko}
\address{Niels Bohr Institute, University of Copenhagen,
Blegdamsvej 17, Copenhagen, 2100-DK, Denmark}
\ead{tuan.pham@nbi.ku.dk}
\vspace{10pt}
\begin{indented}
\item[]
\end{indented}

\begin{abstract}
 The study of adaptive dynamics, involving many degrees of freedom on two separated  timescales, one for fast changes of state variables  and another for the slow adaptation of  parameters controlling  the former's dynamics is crucial for understanding feedback mechanisms underlying   evolution and  learning.  We present a path-integral approach à la Martin–Siggia–Rose-De Dominicis–Janssen (MSRDJ) to analyse nonequilibrium phase transitions in  such dynamical systems. As an illustration, we  apply our framework  to the adaptation of gene-regulatory networks under a 
\emph{dynamic} genotype-phenotype map:
 phenotypic variations are shaped by the fast stochastic gene-expression  dynamics and are coupled to the slowly evolving distribution of genotypes, each  encoded by a network structure. We establish that under this map, genotypes corresponding to reciprocal networks of coherent feedback loops are selected 
within an intermediate range of environmental noise, leading to
 phenotypic robustness. 
\end{abstract}

%
%
%
%
%

\section{Introduction}
In biological or neural systems  parameters that control  the dynamics of the state variables, such as the set of couplings among  the system degrees of freedom (dofs), often  change on a slower timescale compared to that of the dofs' dynamics. Moreover, such  systems become  \emph{adaptive}  if these parameters need to adjust to the long-time  state of the fast dofs. Adaptive multiple-timescale dynamical systems  include   cellular adaptation \cite{Koshland, Stern, Inoue, levine2020phenotypic}, cell differentiation with epigenetic modifications \cite{Waddington,Li2013, Huang2012, Miyamoto, Matsushita2020}, neural networks with synaptic plasticity \cite{gerstner2014neuronal, Duchet2023} or adaptation and synaptic filtering \cite{Beiran,Muscinelli},   sequence retrieval   via interaction modulation \cite{Herron}, eco-evolutionary dynamics \cite{Post, Moran2022, Fraboul}, as well as extensive adaptive network models \cite{Gross2008,BERNER20231}. 
While this class of systems has been analysed  mostly in  low-dimensional models using tools of dynamical systems \cite{Kuehn2015},   dynamical mean-field theory (DMFT) \cite{Hertz2017,Martin, Dominicis,COOLEN2001, Chow_2015}, a powerful tool  to study a large variety of systems with \emph{quenched}   disorder \cite{Sompolinsky_Zippelius_PRL,Sompolinsky, Crisanti1987, Opper1992,Molgedey, Hatchett_2004, Ichinomiya, Galla2005,Stiller, Kadmon2015,Mastrogiuseppe2017,Marti2018,Crisanti2018, Schuecker2018, Agoritsas_2019, Altieri2021, De_Giuli, Baron2023, Pirey, Aguilera2023,behera2023, Clark2023, Pruser2024, shmakov2023coalescence, martorell2023dynamically, garnierbrun2024unlearnable}, does not consider adaptively \emph{slow}  changes in the parameters at all.  Previous works on a related class of systems  with  \emph{partially annealed} disorder and two temperatures, one for the fast dofs and another for  the slow ones, often assume  relaxational dynamics toward thermodynamic equilibrium \cite{Coolen1993,Feldman_1994, Dotsenko_Feldman, Uezu2009, Wemmenhove_2004, Wemmenhove_20041, Rabello_2008, Poderoso2007,Alberici}. High-dimensional nonequilibrium adaptive systems  have challenged these existing approaches and  been  so far much less studied (but see  \cite{Allahverdyan1998} and reference therein for a few exceptions).  

Living systems belong to this class of stochastic adaptive dynamics with two well-separated timescales: 
  phenotypes  evolve under  noise at cellular and molecular levels \cite{Elowitz, Karn,PAULSSON, Eldar2010, Felix, Furusawa2005}, 
while  genotypes -- the rule of such developmental
dynamics -- change  under
selection and mutation on a slower timescale, 
depending on the fitness advantage of   the resulting phenotypes.
A central question in   evolutionary biology  is which 
features of  the genotype-phenotype maps  give rise to
phenotypic robustness against perturbations induced by noise and that against   mutation   \cite{Pham2022,Bressloff_2017, Nichol,Jiang2023, Garcia_Galindo, MANRUBIA2021, KanekoPloSOne2007}. The answer to this question together with the proof of a hypothesis about   an existing correlation between these two different types of robustness under  sufficient  noise, as suggested by  extensive numerical studies   \cite{Bergman, Ciliberti,Crombach, KanekoPloSOne2007, Tadamune, Inoue2021}, nevertheless, remains elusive.  
 A systematic approach is hence required to address this question. Such an approach needs to  bridge  a strong disconnection between the actual nonequilibrium underlying dynamics of development  and their  approximate treatment based on   (quasi)potential landscape picture \cite{Zhou2016, Guillemin}. 


In this  paper,  to account for the \emph{slow} adaptation of the controlling parameters, we develop a so-called adaptive DMFT (ADMFT), thus leveraging the original use of DMFT in quenched-disorder systems to  adaptive ones.  The key point of our approach which is based on the Martin–Siggia–Rose-De Dominicis–Janssen (MSRDJ) formalism \cite{Martin, Dominicis} is to derive effective dynamics for both fast and  slow  variables in the thermodynamic limit 
from 
the  moment generating functional of their joint   trajectories.  
We then apply  our  framework to gene-regulatory networks  with genotype-phenotype feedback. 
Here we find  three phases with regard to the phenotypic  and genotypic states, among which of biological relevance is a region called \emph{robust} phase, where  both genotype and phenotype achieve high values. We show  for  reciprocal networks this robust region  exists   at intermediate noise strength  by describing the  transitions  leading to the emergence of robustness  as the onset of instability of phenotypes with zero gene-expression levels. 

\section{Problem Formulation}
We consider  a system consisting of $N$ units whose states are characterised by a vector $\mathbf{x}$. The  dynamics of  $\mathbf{x}$   take place on a continuous time $t\in[0,\infty)$. Let $\mathbf{J}$ denote the set of parameters controlling the dynamics of $\mathbf{x}$, such as the set of interaction couplings among $\mathbf{x}$'s components. Elements of $\mathbf{J}$ are assumed to be updated  over discrete generations $\tau = 0,1,2,\cdots, T_{\max}$. Let $\mathbf{J}(\tau)$ denote a  configuration of $\mathbf{J}$  at generation $\tau$.  Note that without adaptation, i.e. if there is no dynamics defined over the space of $\mathbf{J}$, one could  interpret $\tau$ as the index of a $\mathbf{J}(\tau)$ configuration in the latter space.  Here   to stress that the fast dynamics of $\mathbf{x}$  depends on which  set $\mathbf{J}(\tau)$ of parameters  that is  currently applied to it,  we denote $\mathbf{x}$'s state by 
   $ x_k(t;\tau) \equiv x_k(t|\mathbf{J}(\tau))$, $ k=1,\cdots, N$, meaning that $x_k$ is an  explicit function  of $t$ only for any given $\tau$. We focus on adaptive dynamics with time-scale separation: 
\begin{itemize}
\item  (i) the fast dynamics of  $\mathbf{x}(t;\tau)$  during which  $\mathbf{J}(\tau)$ is kept \emph{fixed} are assumed to relax toward a non-equilibrium steady state (NESS) with the corresponding distribution  $P_{\mathbf{J}(\tau)}(\mathbf{x})$ as $t\rightarrow \infty$. 
\item (ii) once  $\mathbf{x}$ reaches a NESS, all elements $J_{kj}(\tau)$ of $\mathbf{J}(\tau)$    evolve synchronously  to $J_{kj}(\tau+1)$, each  follows the direction set by a respective field  $ h_{kj}(\tau)$. In our problem, the fields $ h_{kj}(\tau)$  represent  feedback from the steady-state distribution of $\mathbf{x}$  to the adaptation dynamics of $\mathbf{J}(\tau)$. 
Often, these feedback fields are functions of some \emph{fitness}  $\Psi(\tau)$ -- a scalar that contains the information about $\mathbf{x}$  and/or the derivatives of $\Psi(\tau)$ with respect to genetic variables ($J_{kj}$ in this case), i.e., 
\begin{equation}
h_{kj}(\tau) = h_{kj}\big(\Psi(\tau), \nabla_{\bold{J}} \Psi(\tau)\big) \label{h_definition}
\end{equation}
We shall consider  fitness  $\Psi(\tau)$  
 that depends only on the  steady-state distribution  of all the units $P_{\mathbf{J}(\tau)}(\mathbf{x})$ via some (non)linear function $\phi$: 
\begin{equation}
\Psi(\tau) = \int d^{N}x\, P_{\mathbf{J}(\tau)}(\mathbf{x}) \phi(\mathbf{x})\,.
\label{fitness1}
\end{equation} 
\end{itemize}

To make the presentation clear, from now on we specify the elements of $\mathbf{J}(\tau)$ as the interactions among $\mathbf{x}$'s components, e.g. $J_{kj}(\tau)$ is the influence of  $j$ on $k$. We assume that positive and negative influences are the most basic forms of interactions among the system units, so $J_{kj}(\tau)$ can take binary value: $J_{kj}(\tau)=1$ ($J_{kj}(\tau)=-1$) if unit $j$ activates (inhibits) the activity of unit $k$.  This can be ensured for the adaptation of $\mathbf{J}(\tau)$ in (ii) by adopting a discrete-time update with the sign function as shown below. When taking the thermodynamic
limit on fully-connected graphs, proper scaling of $J_{kj}$ will be considered (detailed information on scaling properties of these interactions is provided in Appendix C). In summary, we consider the following unit-coupling dynamics of (i) and (ii) with $h_{kj}(\tau)$ and $\Psi(\tau)$  given in Eqs. \eref{h_definition}-\eref{fitness1}, respective, that are closed and obey:
\begin{equation}
\left \{ \begin{array}{l} \displaystyle 
\left(\frac{ \partial}{\partial t} +1 \right) x_k(t;\tau) =  F\Big(  \sum_{j} J_{kj}(\tau) x_j(t;\tau)   \Big) +\, \xi_k(t; \tau)  \vspace{3pt}\\ 

\displaystyle  J_{kj}(\tau+1)={\rm sign}\,\Big[h_{kj}(\tau)+ \beta^{-1}\tilde{\xi}_{kj}(\tau)\Big]
\end{array} \right. 
 \label{general1}
\end{equation}
where  $J_{kk}(\tau) = 0$, $\forall \tau$, $F(\cdot)$ is a  nonlinear function and $\xi_k(t;\tau)$ is a mean-zero white noise with  $\langle \xi_k(t; \tau) \xi_j(t'; \tau') \rangle =  \sigma^2\delta_{kj}\delta_{\tau \tau'}\delta(t-t')$.  Due to the  non-linearity of $F(\cdot)$ the process $\mathbf{x}(t;\tau)$ in Eq. \eref{general1} does not obey detailed balance (not even for symmetric $\mathbf{J}(\tau)$ \cite{Hatchett_2004}).  
Here a parameter $\beta$ accounts for the strength of stochastic effects in the $\mathbf{J}$'s dynamics induced by a set of `threshold noises' $\{\tilde{\xi}_{kj}\}$.    $\{\tilde{\xi}_{kj}\}$  are independent and identically distributed random variables drawn from a distribution $p(\tilde{\xi})$ that fulfils $p(\tilde{\xi}) =p(-\tilde{\xi})$ \cite{COOLEN2001}. Without the feedback $h_{kj}(\tau)$,   $J_{kj}(\tau +1)$ is a  mean-zero random variable due to the term $\beta^{-1}\tilde{\xi}_{kj}(\tau)$.

We shall implement the dynamics of Eq. \eref{general1} in a nested fashion. Specifically, at generation $\tau$, we integrate the first equation  under the present interaction matrix  $\mathbf{J}(\tau)$ until it reaches a steady-state solution  $\mathbf{x}(\infty;\tau)$, and then we use this $\mathbf{x}(\infty;\tau)$ to compute the feedback fields $h_{kj}(\tau)$ via Eqs. \eref{h_definition}--\eref{fitness1}; once all the $h_{kj}(\tau)$ are known,  we update $\mathbf{J}(\tau)$ to $\mathbf{J}(\tau+1)$ according to the second equation of Eq. \eref{general1}. This procedure is again repeated at generation $\tau+1$. Underlying this implementation of  Eq. \eref{general1} is our  assumption that the duration of one generation must be long enough  for the subsystem of  fast variables to relax to its asymptotic attractor, which is a steady state defined by the interaction matrix at that generation. This is indeed the main assumption  of many \emph{genetic algorithms}  \cite{Bergman, Ciliberti,Crombach, KanekoPloSOne2007, Tadamune, Inoue2021} and has also been used recently in  the evolutionary dynamics of ecological communities \cite{Fraboul}. 

\section{The evolution of genotype-phenotype map}
\subsection{Model description}
  We now describe 
  a model of genotype-phenotype maps in  \cite{KanekoPloSOne2007} that represents the genotype by $\mathbf{J}$ and the phenotype  by   $\mathbf{x}$.  In this model  $\mathbf{J}$ is  a gene regulatory network that has an all-to-all topology, while  $\mathbf{x}$ are the gene expression levels of $N$ genes on this network. 
Motivated by the observation that   the functionality of genetic networks is often determined only by a fraction of genes, the model introduces 
 the concept of \emph{target} genes (units). Their set is denoted by $\mathcal{T}$ and their number -- by  $N_t$. All other genes   are called non-target, their set is denoted by $\mathcal{O}$ and their number -- by $N_o = N- N_t$. A fixed ratio of target to non-target is considered in the thermodynamic limit:
\begin{equation}
\alpha=N_o/N_t = O(1)\,,\quad   N\rightarrow \infty
\label{alpha}
\end{equation}  
The model  assumes that non-target genes do not contribute directly to the   \emph{absolute} fitness $\Psi(\mathbf{J}(\tau))$,  i.e., for a given network configuration $\mathbf{J}(\tau)$, 
$\phi(\mathbf{x})$ is   just a function of $x_i$, for $i\in \mathcal{T}$: 
\begin{equation}
\Psi(\tau) \equiv \Psi(\mathbf{J}(\tau)) := N_t^{-2}\int d^{N_t}x\, P_{\mathbf{J}(\tau)}(\mathbf{x}_{\mathcal{T}})\Big(\sum_{i\in \mathcal{T}}   x_i\Big)^2\,,
\label{fitness}
\end{equation}
where $P_{\mathbf{J}(\tau)}(\mathbf{x}_{\mathcal{T}}):=\int d^{N_o}x P_{\mathbf{J}(\tau)}(\mathbf{x})$ is the marginal distribution of target genes.  This definition of fitness reflects the fact that the overall expression of some target (important) genes is a good proxy for the organism's chance of    survival and reproduction. In the context of neural networks with learning, such as FORCE learning \cite{SUSSILLO2009544}, the readout neurons play the role of target units as their states determine the network performance in terms of some loss functions. For eco-evolutionary systems, target units represent the keystone (focal) species, i.e. those that have strong impacts on the communities  in which they reside  \cite{Post}.  
The partition into the target and non-target genes results in 4 different types of interactions, namely,   $J_{ij}^{(tt)}$ for $i \in \mathcal{T}$ and $j \in \mathcal{T}$;   $J_{ij}^{(oo)}$ for $i  \in \mathcal{O}$ and  $j \in \mathcal{O}$;  $J_{ij}^{(to)}$ for $i \in \mathcal{T}$ and $j \in \mathcal{O}$; $J_{ij}^{(ot)}$ for $i \in \mathcal{O}$ and $j \in \mathcal{T}$.  A similar division into two groups has been formulated for spin-systems, where,  under certain conditions, target spins are in an effective equilibrium, while non-target spins remain non-equilibrium \cite{Saad2013}. 

By excluding the terms with $k=j$ from   the definition of $\Psi(\tau)$ in Eq. \eref{fitness} as they are only subleading correction of order $O(N_t^{-1})$ as $N_t \rightarrow \infty$, we arrive at:  
\begin{equation}
 \Psi(\tau) = \frac{1}{N_t^2} \sum_{k\neq j \in \mathcal{T}} \big[x_kx_j\big]_{\tau} \,,\quad \big[x_kx_j\big]_{\tau} := \int d^{N_t}x\, P_{\mathbf{J}(\tau)}\big(\mathbf{x}_{\mathcal{T}}\big) x_kx_j \label{squared_fitness}
\end{equation}
\subsection{Specification of the  feedback fields in \Eref{h_definition}}
 As $h_{kj}(\tau)$ 
 depends on which groups ($\mathcal{T}$ or $\mathcal{O}$) that $i$ and $j$ belong to, 
 to distinguish  the feedback fields that act on  $J_{ij}^{(tt)}$, $J_{ij}^{(oo)}$, $J_{ij}^{(to)}$, $J_{ij}^{(ot)}$, we denote them by  $h_{ij}^{(tt)}$,  $h_{ij}^{(oo)}$, $h_{ij}^{(to)}$ and  $h_{ij}^{(ot)}$, respectively. Here we  provide a  specification of  Eq. \eref{h_definition} that makes use of two facts: (i) since the fitness $\Psi(\tau)$ in Eq. \eref{squared_fitness} depends \emph{directly}  on the target genes only, its increments should coincide (on average) with those of  $h_{ij}^{(tt)}(\tau)$ and (ii) its gradient  with respect to the intergroup couplings $J_{k\ell}^{(to)}(\tau) $ should be the force driving the 
evolution of $h^{(to)}_{k\ell}(\tau)$.
By assuming that these facts can be  written as follows
\numparts 
\begin{eqnarray}
 \sum_{i\neq j \in \mathcal{T}}\Big(h_{ij}^{(tt)} (\tau+1)- h_{ij}^{(tt)} (\tau)\Big) = N_t^2\big[\Psi(\tau+1) - \Psi(\tau) \big]\,,
\label{h_tt_as_average_fitness}\\
 h_{k\ell}^{(to)}(\tau+1) - h_{k\ell}^{(to)}(\tau)  =\frac{\partial \Psi}{\partial J_{k\ell}^{(to)}(\tau) }\,,\qquad k\in\mathcal{T}\,, \ell\in\mathcal{O}\,,
 \label{fitness_gradient}
\end{eqnarray}
\endnumparts 
then we can  introduce [see  Appendix B for the details]:
\begin{equation}
 h_{ij}^{(tt)} (\tau)  :=  \big[x_ix_j\big]_{\tau}\,,\quad (i,j) \in \mathcal{T}\,.
\label{h_for_Jtt} 
\end{equation}
\begin{equation}  h_{k\ell}^{(to)}(\tau)  :=  N_t^{-2}\Big(1- \big[x_k^2\big]_{\tau}\Big) \sum_{i\neq j\in \mathcal{T}} J_{i\ell}^{(to)}(\tau)
 J_{\ell j}^{(ot)}(\tau)\big[x_ix_j\big]_\tau\,. \label{h_for_Jto}
 \end{equation}
These equations indeed ensure  the system follows the direction of increasing fitness. In particular, by  expressing $h_{ij}^{(tt)}$ as phenotypic covariances in
Eq. \eref{h_for_Jtt} -- also known as the Hebbian  rule \cite{hebb1949}, we can describe the adaptation of intragroup couplings as a Hopfield model with evolving patterns, similar to \cite{Schnaack}. For  large $\beta$ in Eq. \eref{general1}, this learning rule results in symmetric interactions among target genes, i.e.  $J^{(tt)}_{kj}(\tau) =  J^{(tt)}_{jk}(\tau)$,  $\forall \, \tau$. Finally, while   Eq. \eref{h_for_Jto} is chosen to satisfy  the relation between changes of $h_{k\ell}^{(to)}(\tau)$ and the  gradient of the fitness, this form admits an intuitive biological interpretation in terms of how coherent feed-forward motifs, i.e. $J_{i\ell}^{(to)} J_{\ell j}^{(ot)}>0$,  contribute to a fitness increment.

 In a population of individuals with different networks (genotypes),  the adaptation step of \cite{KanekoPloSOne2007} follows a genetic algorithm that,  at each generation $\tau$,  keeps only those  individuals whose \emph{relative} fitness $\Psi(\tau)/\langle \Psi(\tau)\rangle_{\mathbf{J}(\tau)}$ higher than some threshold  
 for the production of offsprings at generation $\tau+1$, according to Fisher's  theorem \cite{Fisher} [here $\langle \cdot\rangle_{\mathbf{J}(\tau)}$ denote averaging taken wrt the distribution $\tilde{P}(\mathbf{J}(\tau))$  over the ensemble of  networks at generation $\tau$]. 
In accordance to this algorithmic use of the relative fitness, we  introduce a scaling $\Psi(\tau) \rightarrow \Psi(\tau)/\langle \Psi(\tau)\rangle_{\mathbf{J}(\tau)}$ into Eqs. \eref{h_for_Jtt}-\eref{h_for_Jto}, leading to 
\numparts 
\begin{eqnarray}
 h_{ij}^{(tt)} (\tau)  :=  \frac{ \big[x_ix_j\big]_{\tau}}{\langle \Psi(\tau)\rangle_{\mathbf{J}(\tau)}}\,,
\label{h_for_Jtt_relative}\\
 h_{k\ell}^{(to)}(\tau)  :=  N_t^{-2}\left(1-\frac{\big[x_k^2\big]_{\tau}}{\langle \Psi(\tau)\rangle_{\mathbf{J}(\tau)}}\right)  \sum_{i\neq j\in \mathcal{T}} J_{k\ell}^{(to)}(\tau)
 J_{\ell j}^{(ot)}(\tau) \frac{ \big[x_ix_j\big]_{\tau}}{\langle \Psi(\tau)\rangle_{\mathbf{J}(\tau)}}\,.\qquad
  \label{h_for_Jto_relative}
\end{eqnarray}
\endnumparts 
In summary, we formulated the adaptive dynamics of the model \cite{KanekoPloSOne2007}  in the form of Eq.  \eref{general1}, where   the gene-expression dynamics of $\mathbf{x}$ correspond to a choice of $F(\cdot)= {\rm tanh}(\cdot)$: 
\begin{equation}  \left(\frac{\partial}{\partial t} +1\right)  x_k(t;\tau) = {\rm tanh}\Big(\sum_{j} J_{kj}(\tau) x_j(t;\tau) \Big)  + \xi_k(t;\tau)\,,\label{GRN_dynamics}
\end{equation} 
and the adaptation process of $\mathbf{J}$ is specified by  $h_{kj}^{(tt)}(\tau)$  and $h_{k\ell}^{(to)}(\tau)$  in Eqs. \eref{h_for_Jtt_relative}-\eref{h_for_Jto_relative}, using the fitness  defined in Eq. \eref{squared_fitness}.


\subsection{Adaptive Dynamical Mean-Field Theory (ADMFT)}
In this section, we present an analytical framework for studying the model introduced in the previous section.  As we consider  fully-connected networks, for the thermodynamic limit $N_t\rightarrow \infty$,  the couplings $ J^{(tt)}_{kj}$ should be rescaled by $1/\sqrt{N_t}$ to ensure a sensible thermodynamic limit. We define the averaged coupling among the target genes $\hat{\mu}(\tau)$ as
\begin{equation}
   \hat{\mu}(\tau):= \frac{1}{N_t(N_t-1)}\sum_{(i,j)\in\mathcal{T}}  J^{(tt)}_{ij}(\tau) \label{jtt}\,.
\end{equation}
In Appendix C, we derive 
\begin{equation}
  \hat{\mu}(\tau+1)  = {\rm tanh}\big( \beta h^{(tt)}(\tau) \big) \label{jtt_dynamics}
\end{equation} 
Moreover, it is necessary  to quantify  the symmetry level of the couplings by a parameter $\nu$ that, for any pair of  $i $ and $ j $, is equal to   the  covariance between $J_{ij}$ and $ J_{ji}$\footnote{This average  is taken wrt  the distribution $\tilde{P}(\mathbf{J}(\tau))$  over the ensemble of networks at generation $\tau$.}:
\begin{equation}
\nu:=N_t \Big\langle   \big[J_{ij} - \langle J_{ij} \rangle \big]\big[J_{ji} - \langle J_{ji}\rangle\big] \Big\rangle_{\mathbf{J}(\tau)}
\end{equation}
In general, $\nu\in[-1,1]$. In particular, the specific cases $\nu= \{1, 0, -1\}$ correspond to  fully symmetric, asymmetric  and   
and  antisymmetric  interactions. To simplify the analytical treatment we fix $\nu$ as a generation-independent constant, i.e. $\nu(\tau)= \nu(\tau+1) = \nu$, $\forall\, \tau$, by imposing appropriate constraints on the relation between $h_{ij}(\tau)$ and $h_{ji}(\tau)$. 

Hereafter we only focus on the dynamics of the target genes. For notational simplicity, we drop $\mathcal{T}$ from $\mathbf{x}_{\mathcal{T}}$. So $\mathbf{x} = \big(x_1,x_2,\cdots, x_{N_t}\big)$.   While  binary couplings are used in Eq. \eref{general1},  in deriving the ADMFT, for $(i,j)\in \mathcal{T}$, we shall replace  $J^{(tt)}_{ij}(\tau)$ by the effective couplings $J^{(e)}_{ij}$ given in Eq. \eref{signal_to_noise} of Appendix A.  $J^{(e)}_{ij}$  have continuous values and  variance equal to $\alpha/N_t$. Using the path integral formalism \cite{Martin, Dominicis,COOLEN2001, Chow_2015, Hertz2017} as detailed in Appendix C, 
we can derive from  a saddle-point approximation that becomes exact as $N\rightarrow \infty$, an effective process of a single unit  $x(t; \tau)$  whose distribution of trajectories is 
\numparts 
\begin{eqnarray}
\mathcal{P}\big(\{x\}
    \big|\{\eta\}, \{\xi\}\big) =\prod_{\tau=0}^{T_{\rm max}-1} \int dt\Big\langle \delta\big(\partial_t x + x - F(\kappa)  - \xi \big) \Big\rangle_* \label{closed0} \\
   \kappa(t; \tau)=\hat{\mu}(\tau) m(t; \tau)  +   \alpha\nu \int_0^t dt' G(t,t'; \tau)x(t'; \tau) +\eta(t; \tau)
   \label{kappa}
\end{eqnarray}
\endnumparts 
 where $\langle \cdot\rangle_*$ denotes the average taken wrt this effective measure $    \mathcal{P}\big(\{x\}|\{\eta\},\{\xi\}\big)$. The averaged activity of this variable $m(t; \tau)$, its autocorrelation  $C(t,t';\tau)$, its response  function $G(t,t';\tau)$ and the effective noise $\eta(t'; \tau)$, all are  self-consistently defined  from:
 \numparts 
 \begin{eqnarray}
m(t;\tau) &:=& \big\langle x(t;\tau)\big\rangle_*  
\label{orderparameter1}
\\ C(t,t';\tau) &:=&\big \langle x(t;\tau) x(t';\tau)\big\rangle_*  \qquad\label{orderparameter2}\\
    G(t,t';\tau) &:=&\left \langle \frac{\delta x(t;\tau)}{\delta \eta(t';\tau)}\right\rangle_*\,,\label{orderparameter3} \\ \big\langle \eta(t;\tau) \eta(t'; \tau) \big\rangle_* &:=&  \alpha\, C(t,t'; \tau) 
\label{self_consistency_noise}
\end{eqnarray}
\endnumparts 
 When $\nu = 0$,   the term associated with the response function drops out, leading to dynamics similar to that of \emph{static} asymmetric networks \cite{martorell2023dynamically}. If  $\alpha=0$, the system becomes  effectively equilibrium. Once integrating the effective process  associated to the measure in Eq. \eref{closed0}  into the  evolution of $  \hat{\mu}(\tau)$ in Eq. \eref{jtt_dynamics},  we arrive at:  
\begin{equation}
\left \{ \begin{array}{l} \displaystyle  \hat{\mu}(\tau+1)  = {\rm tanh}\big( \beta h(\tau) \big)  \vspace{3pt}\\  \displaystyle  \partial_t x(t; \tau) = -x(t; \tau)  + F(\kappa(t;\tau)) +\, \xi(t; \tau) 
     \end{array} \right.\, 
  \label{closed1}
 \end{equation}
 In the $N\rightarrow \infty$ limit, the  \emph{unit-coupling} dynamics of Eq.   \eref{closed1} yield the same statistics as those of the  dynamics in Eq. \eref{general1} for the model \cite{KanekoPloSOne2007}, in an analogous manner to what was rigorously proven for random neural networks \cite{Cabana}. 
 Numerical  solutions to the second of  Eq.   \eref{closed1}  
can be found using Monte Carlo methods  \cite{Eissfeller, Roy2019, zou2023,Breffle2023}. Once  its asymptotic attractors $x(\infty; \tau)$ have been found, we   update  $\hat{\mu}(\tau)$ using the first of  Eq.   \eref{closed1}, provided that the form of $h(\tau)$  is specified. This  set of closed dynamical equations in  Eq.   \eref{closed1}  then can be iterated over many generations $\tau$, in a nested manner, i.e. in the same way as the aforementioned implementation of the full dynamics Eq. \eref{general1}. 
%


\subsection{Fixed-point solutions for fully symmetric couplings}
In this section we use the  ADMFT framework to quantify
the system behaviour at steady state by its order parameters. 
Assuming that at generation $\tau$ the target unit  $x$'s  dynamics reach  a stationary state with time-translational symmetry,  $\forall{s}>0$, we have
\numparts 
\begin{eqnarray}
\lim_{t'\rightarrow \infty}C(t=t'+s,t', \tau) = C(s, \tau)  \\ \lim_{t'\rightarrow \infty}G(t=t'+s,t', \tau) =  G(s, \tau)  \\ \lim_{t\rightarrow \infty}G(t,t', \tau) =0\,,\forall\, (t',\tau)
\end{eqnarray}
\endnumparts 
As well-known from previous studies on static fully-connected  networks \cite{Crisanti2018},  a large value of $\alpha$ would drive the system towards chaotic attractor, thus breaking this assumption that the fast-time
dynamics converge to a well-defined stationary state. We will elaborate on the restriction of $\alpha\leq 0.5$  for a fully-connected  system to reach a stable fixed-point solutions  in Appendix E.
 We define the  
 integrated response $\hat{\chi}$ (that should remain finite for fixed-point solutions)  as follows 
\begin{equation}
 \hat{\chi}(\tau) = \int_0^{\infty}  ds \,G(s, \tau)     
 \end{equation}
   When  $x(t; \tau)$ fluctuates around a $t$-independent average at a given generation $\tau$, the  steady-state autocorrelation $C(s, \tau)$ reaches a  positive (possibly) plateau value as $s\rightarrow \infty$: $\hat{q}(\tau) :=\lim_{s\rightarrow\infty} C(s, \tau)$. After taking this limit, we take the limit $\tau \rightarrow \infty$ to identify 
 fixed points of the unit-coupling system Eq. \eref{closed1} that are denoted by 
\begin{eqnarray}
\mu = \lim_{\tau\rightarrow \infty} \hat{\mu}(\tau)\,, \quad  x_* = \lim_{\tau, t \rightarrow \infty} x(t, \tau)\,,\quad  &\chi = \lim_{\tau\rightarrow \infty} \hat{\chi}(\tau)\,,\quad  q =\lim_{\tau\rightarrow \infty} \hat{q}(\tau)\,
\end{eqnarray}
 Fixed points of Eq. \eref{closed1} then satisfy:
\begin{equation}
    x_* = {\rm tanh}\left(\mu \langle x_*\rangle_{\eta, \xi} + \alpha\chi  x_*  \,+\eta \right) + \xi\,
\label{fixed_point_of_closed2}
\end{equation}
This equation, for  given  distributions of $\eta$ and $\xi$, defines the distribution of $x_*$.
As $x$ reaches a stationary state, so does $\eta$: $\lim_{t,\tau\rightarrow \infty} \eta(t,\tau) = \eta_0= J_0\sqrt{q}\tilde{z}$, where  $\eta_0$ is a Gaussian  random number with  mean zero and variance $ J_0^2 q$, $J_0 =\sqrt{\alpha}$ and  $\tilde{z} \sim\mathcal{N}(0,1)$, the standard normal distribution. Likewise, at stationary, the effect of the white noise $\xi$ is equivalent to  a static  random number $\xi_0 = \sigma z$, 
for $z \sim\mathcal{N}(0,1)$. Taking these facts into account, we can rewrite Eq. \eref{fixed_point_of_closed2} as
\begin{equation}
    x_*(\tilde{z},z) = {\rm tanh}\left(\mu m_\infty  + \alpha\chi  x_*  \,+\,J_0\sqrt{q}\tilde{z} \right) + \sigma z\,,\qquad m_\infty = \langle x_*\rangle_{z,\tilde{z}}\,.
\label{fixed_point_of_closed1}
\end{equation}
In general, Eq. \eref{fixed_point_of_closed1}
can have multiple solutions, depending on initial conditions \cite{Breffle2023}. We, however,  assume that  $ x_*(\tilde{z},z)$ is unique for any given realisation of $\eta$ and $\xi$.  
This allows us to obtain  the following set of self-consistency equations for the averages taken over the ensemble of fixed points characterising the genotype-phenotype stationary state:  
\begin{equation}
 \left \{ \begin{array}{l} \displaystyle
 \mu = {\rm tanh} \big(\beta m^2_\infty/q\big)  \vspace{4pt} \\  m_\infty= \int_{-\infty}^{\infty}  Dz   \int_{-\infty}^{\infty} D\tilde{z}
\,x_*(\tilde{z},z) \vspace{4pt}\\ \displaystyle
   q = \int_{-\infty}^{\infty} Dz \int_{-\infty}^{\infty} D\tilde{z} \big( x_*(\tilde{z},z)\big)^2  \vspace{4pt} \\
   \chi = \displaystyle \int_{-\infty}^{\infty}  Dz  \int_{-\infty}^{\infty}  D\tilde{z} \, \frac{1-f_0^2}{1-\alpha\chi\big(1-f_0^2\big) } 
  \end{array} \right.\,    \label{closed2}
\end{equation}
where  $f_0 := {\rm tanh}\left(\mu m_\infty  + \alpha\chi  x_*  \,+\,J_0\sqrt{q}\tilde{z} \right)$ and $Dz := dz e^{-z^2/2}/\sqrt{2\pi}$ is the Gaussian measure; $m_\infty$ is the average  of  $x_*$; $q$ is the average of $x^2_*$ ;  $ \chi$ is the integrated response of $x_*$; $\mu$ is  the steady-state mean value of $J^{(tt)}$. Solutions are obtained as stable attractors of  iterative dynamics started from initial conditions  with  sufficiently large $m_\infty$, $q$,  $\mu$. To separate the effect of the noise $\xi$  on the system behaviour from that of the self-sustained fluctuations $\eta$, instead of $q$, hereafter we consider the intrinsic variance  $q_0$ defined as:  
\begin{equation}
    q_0= q - \sigma^2\,.
\end{equation} 

\subsection{Spectral indicator of the genotypic critical transitions}

Here to quantify the functional relationships between target and non-target genes that emerge through the adaptive dynamics, we  consider  the case of symmetric intergroup couplings, i.e., for any pair of  $k \in \mathcal{T}$ and $ j \in \mathcal{O}$,
 $J_{kj}^{(to)} = J_{jk}^{(ot)}$. We then define the averaged couplings  between target and non-target genes as
\begin{equation}
   \hat{\lambda}(\tau) := \frac{1}{N_t N_o}\sum_{i \in \mathcal{T},j \in \mathcal{O}}  J^{(to)}_{ij}(\tau)  := \frac{1}{N_t N_o}\sum_{i \in \mathcal{T},j \in \mathcal{O}}  J^{(ot)}_{ji}(\tau)  
\label{jto}
 \end{equation}
 for which one can show that $\hat{\lambda}(\tau+1) = {\rm tanh}\big( \beta h^{(to)}(\tau) \big) $, as given in Appendix C.
The steady-state average  of $J^{(to)}$ (and  $J^{(ot)}$) can be computed from [see
Appendix B for the derivation]:
\begin{equation}
\lambda :=\lim_{\tau\rightarrow \infty} \hat{\lambda}(\tau) = {\rm tanh} \left[\beta \lambda^2 \frac{m^2_\infty}{q} \left(1- \frac{m^2_\infty}{q}\right)\right] 
\label{steady_state_jto}
\end{equation}
Note  $\lambda =0$ is always a solution of this equation.   We are  particularly interested in the critical transition between this \emph{trivial} solution  and the \emph{non-trivial} ones that can only emerge if $m_\infty\neq 0$. 
 It is thus important to know when the system relaxes to a steady state with non-zero activity, $m_\infty\neq 0$. To this end, we need to analyse the local stability of $\mathbf{x} = \mathbf{0}$ (i.e. $x_i=0$, $\forall \, i\in \mathcal{T}$) by using   the largest eigenvalue $\Lambda_1$ of the Jacobian matrix at  $\mathbf{x} = \mathbf{0}$ (i.e. by analysing the linearised subsystem of target units). We   obtain $\Lambda_1$ from $\mu$ and $\lambda$ (see Appendix D for the  derivation):
\begin{equation}
    \Lambda_1 = \frac{\lambda^4 +\mu^2}{\mu} - 2 \label{largesteigenvalue}
\end{equation}
 This equation shows how $\Lambda_1$ depends on the  steady-state behavior of the genotypic variables $\mu$ and $\lambda$:  $\Lambda_1 =-2$ if $\mu=\lambda=0$, $\Lambda_1 =-1$ if   $\lambda=0$ but $\mu=1$ and $\Lambda_1 =0$ if   $\mu=\lambda=1$. Therefore, apart from being a measure of stability, $\Lambda_1$ is also a good indicator of the genotype's steady states.

\section{Results}
 We now present results for $\alpha=0.5$, similar behaviour is observed for other $\alpha > 0.07$ (see Fig. 2 \textbf{(B)} below, where $\alpha =0.07$ is the minimal value of $\alpha$ for which solutions with non-zero $\lambda$ can be found). The particular value of $\alpha =0.5$ is chosen for the sake of presentation only as in this case one can find a robust region over the  largest window of noise.
 The steady-state behaviors and, in particular, the emergence of robustness  can be quantified by the dependence of  $\Lambda_1$ computed from Eq. \eref{largesteigenvalue} on $T_J = 1/\beta$ and $\sigma$. 
 
\begin{center}
\begin{figure}[t]
\includegraphics[scale=0.25]{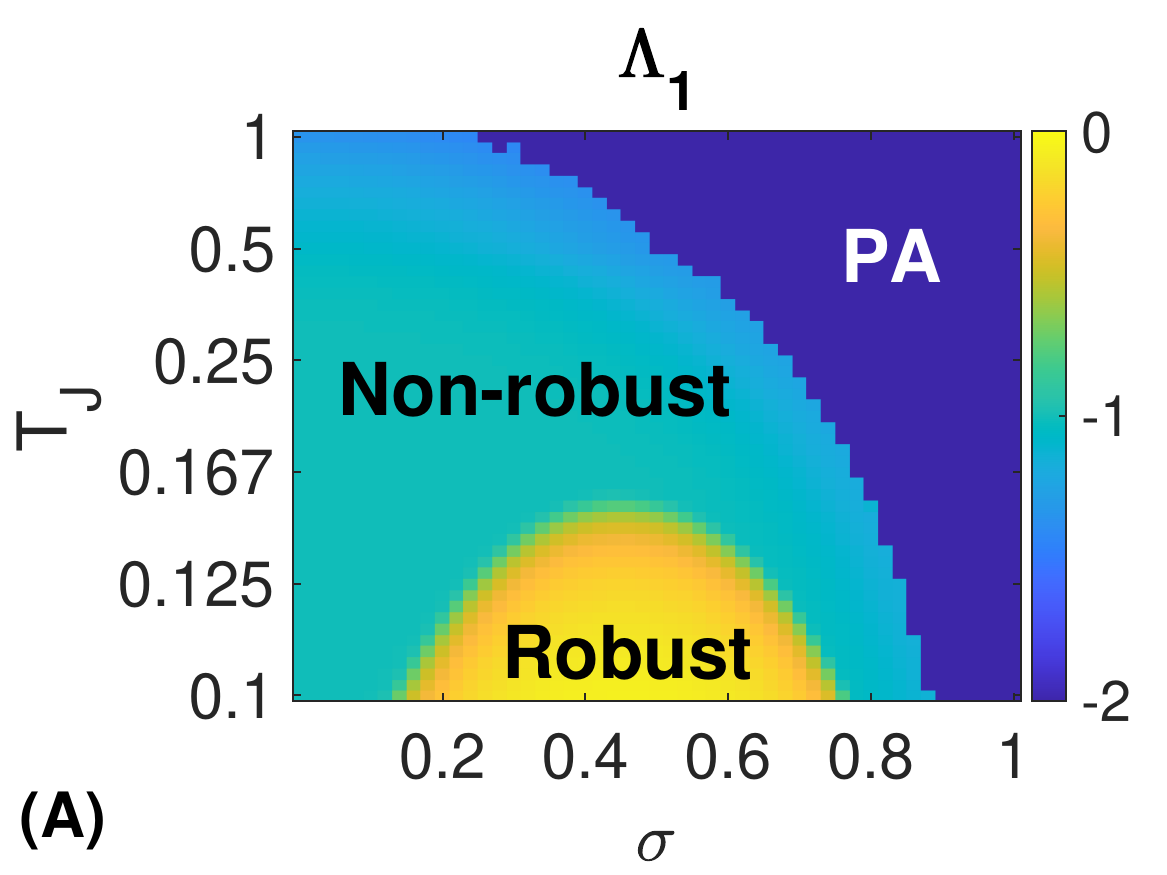}
\includegraphics[scale=0.25]{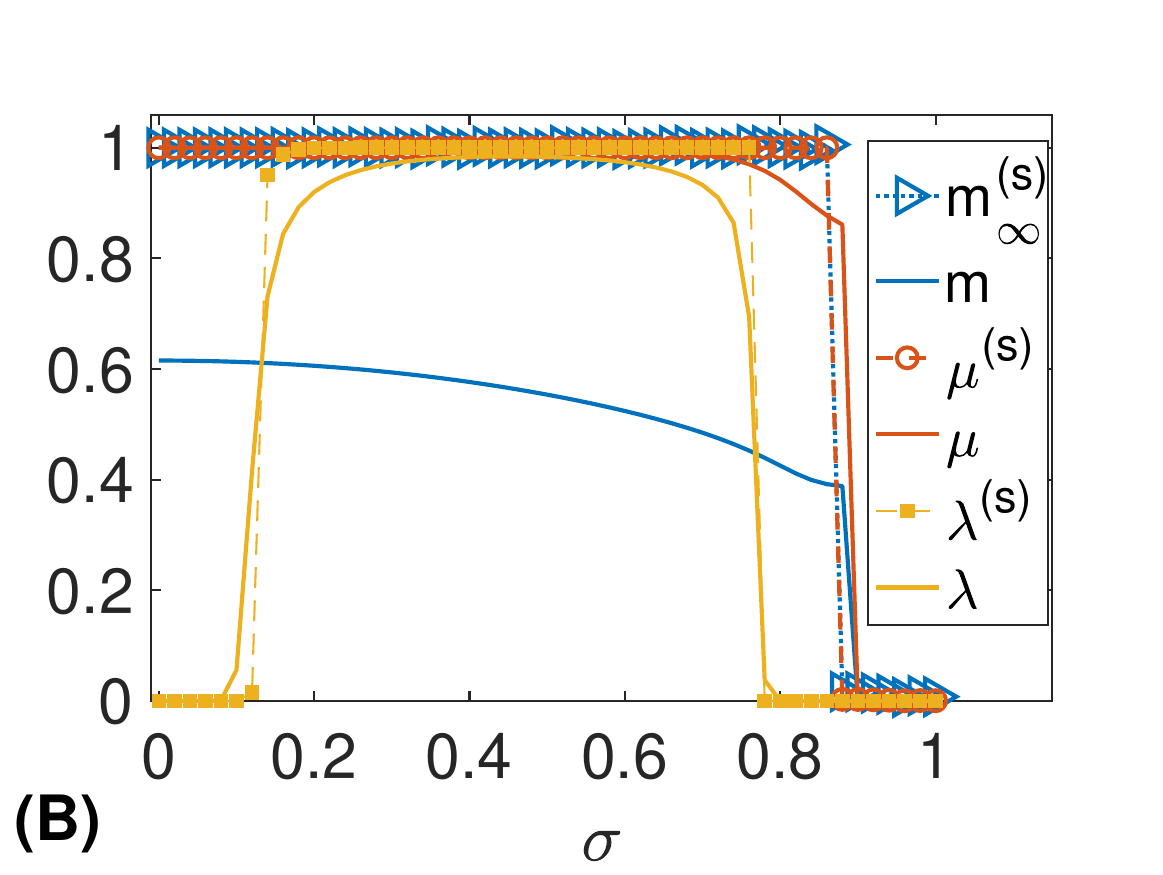}
\includegraphics[scale=0.25]{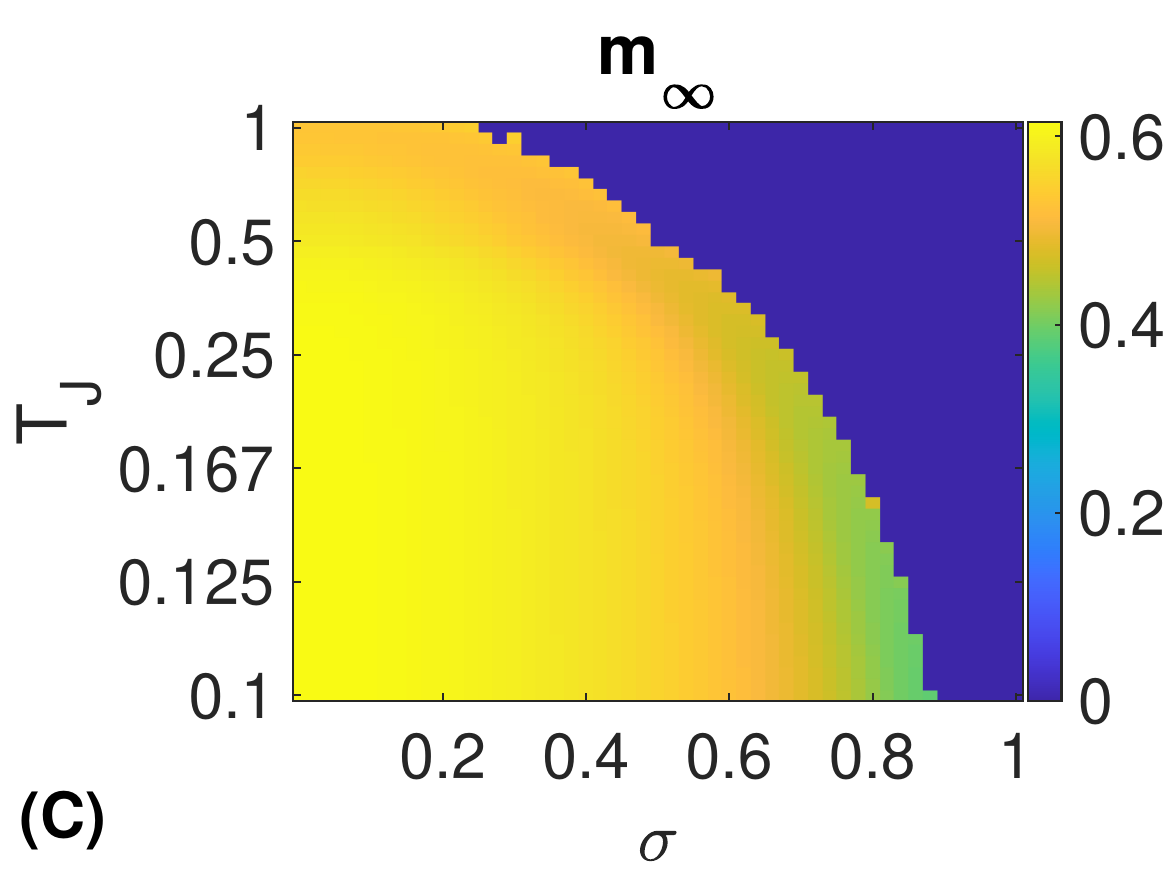}

\includegraphics[scale=0.25]{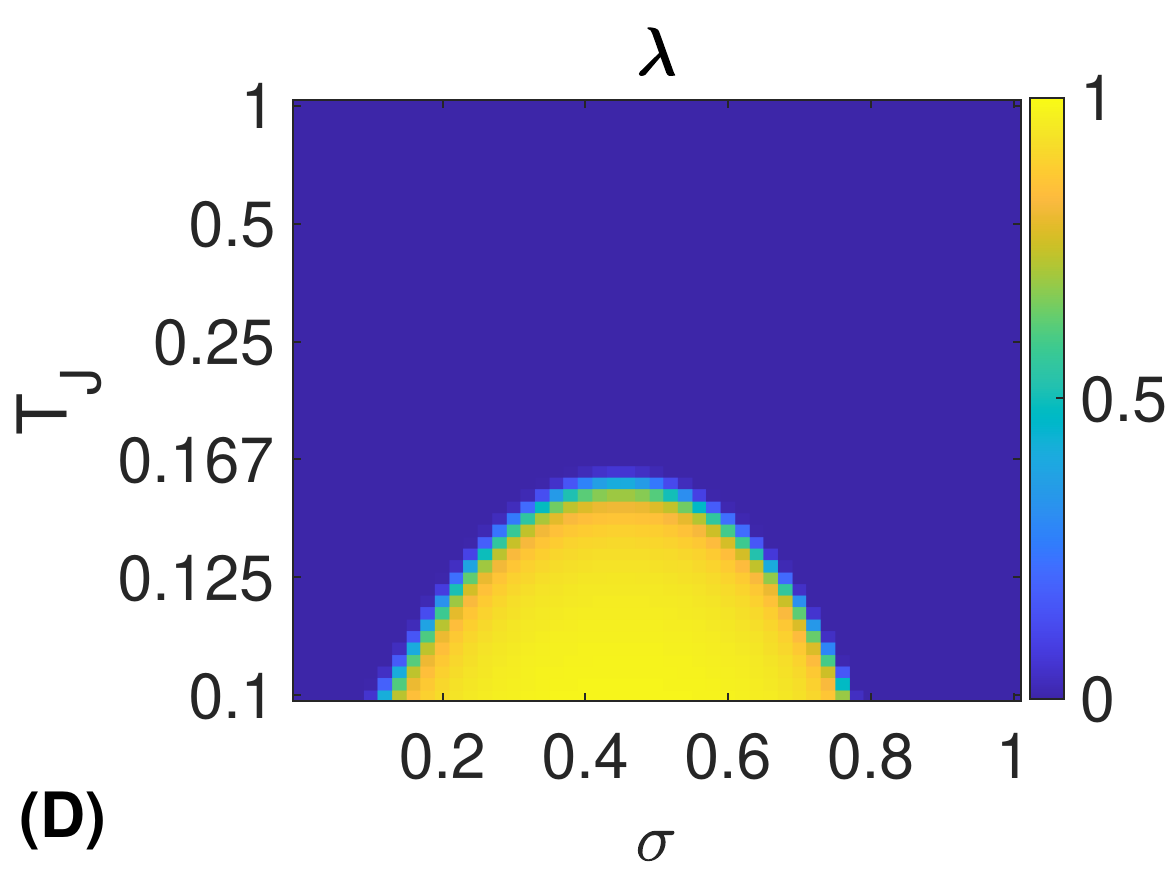}
\includegraphics[scale=0.25]{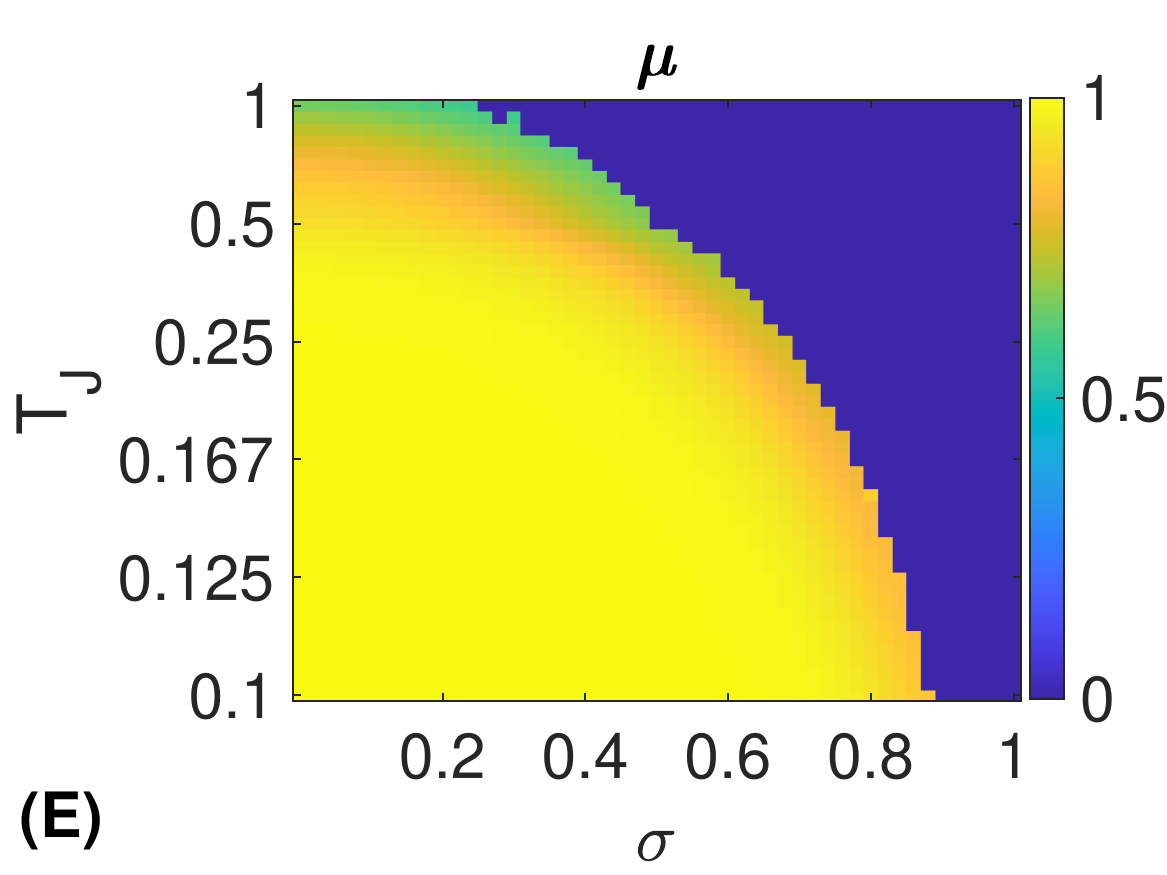}
\includegraphics[scale=0.25]{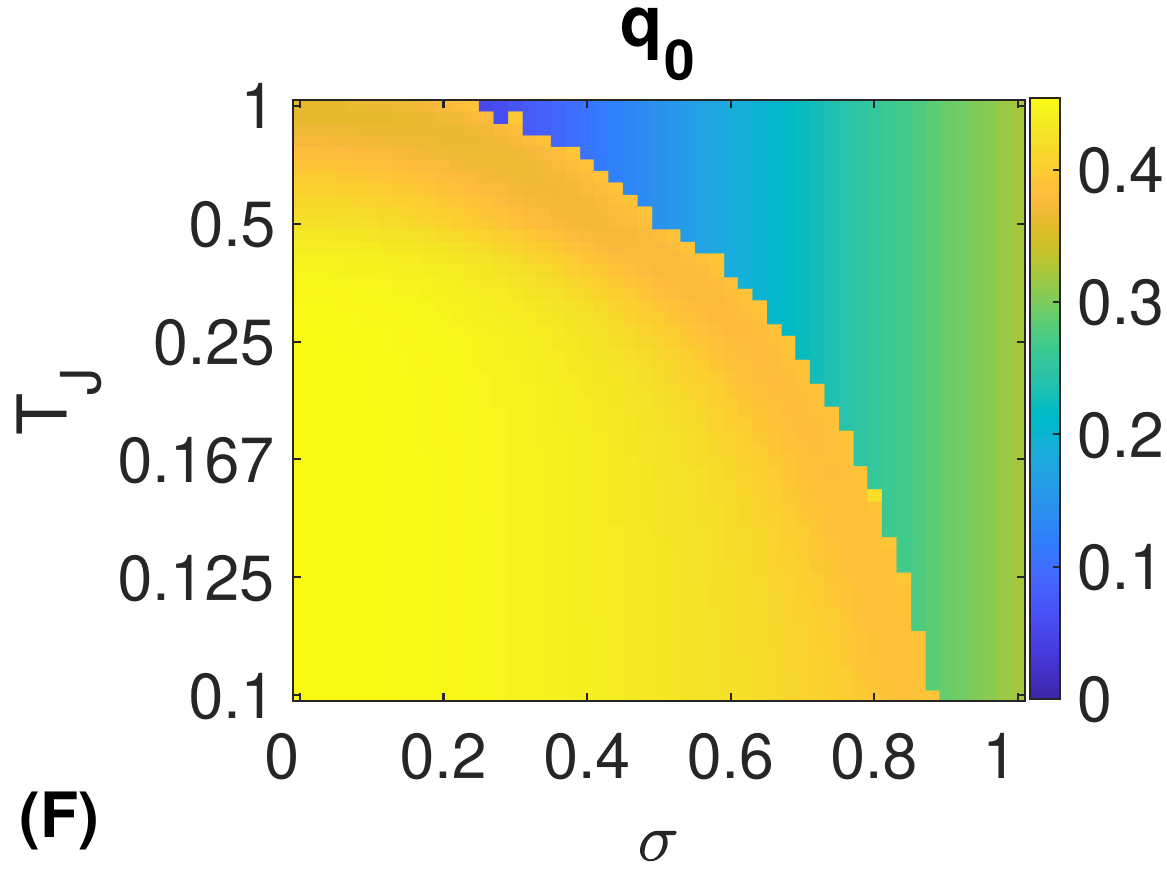}
\caption{\textbf{(A)} The largest eigenvalue $\Lambda_1$ as function of $\sigma$ and $T_J =\beta^{-1}$. Here $\Lambda_1$ is computed by Eq. \eref{largesteigenvalue} for the linearised subsystem of target genes at $\mathbf{x} = \mathbf{0}$. Three distinct values of $\Lambda_1$  indicate three different phases that are called robust, non-robust, and para-attractor as explained in Table 1. \textbf{(B)} Order parameters ($m_\infty$, $\lambda$ and $\mu$) as function of $\sigma$ for $\beta=10$. In the non-robust phase $m_\infty, \mu >  0$  but $\lambda =  0$, while in the robust phase    $m_\infty , \lambda, \mu >0$. In the para-attractor phase \textbf{(PA)}   $m_\infty = \lambda = \mu=0$. These order parameters are obtained as solutions to  Eqs. \eref{closed2}-\eref{steady_state_jto} and are depicted by continuous lines.  Here we also include results from numerical simulations ($m^{(s)}_\infty$, $\lambda^{(s)}$ and $\mu^{(s)}$) of the original adaptive dynamics in Eq. \eref{general1} and depict them by symbols.  \textbf{(C)} Averaged activity of target genes $m_\infty$.  \textbf{(D)}  The average of target vs non-target coupling $\lambda$.  \textbf{(E)}   The  average of target vs target coupling $\mu$. \textbf{(F)} The intrinsic variance in the  target gene's activity  $q_0$.   In all panels $\alpha=0.5$ and $\nu=1$. For simulations in \textbf{(B)}   $N=300$ is used. } 
\label{fig:fig1}
\end{figure}
\end{center}

 In  Fig. \ref{fig:fig1} \textbf{(A)}  we find  three distinct regions corresponding to different values of  $\Lambda_1$, namely,  $\Lambda_1=-2$, $\Lambda_1\simeq -1$ and  $\Lambda_1$ close to zero. This is because of a non-monotonic dependence of $\lambda$ on $\sigma$ at  high enough selection pressure $\beta$. Specifically, below some $T_J^{(\lambda)}$, upon increasing $\sigma$, $\lambda$ first increases from zero to a plateau value close to 1 and then drops to zero again. Such behavior is demonstrated  for $\beta = 10$ in   \fref{fig:fig1}  \textbf{(B)}, confirming three possible solutions of Eq. \eref{closed2} and Eq. \eref{steady_state_jto}: the first solution is $\lambda =0$ and $m_\infty, \mu>0$, the second  corresponds to $m_\infty, \lambda, \mu >0$ and the last  -- to  $m_\infty= \lambda = \mu =0$. We call them non-robust, robust and para-attractor, respectively. Their characteristics are summarised in Table 1.  Note that due to a positive feedback between  $\mu$  and $m_\infty$ that is imposed by the first of Eq. \eref{closed2},  $m_\infty$ and $\mu$ always behave similarly. As expected, we find the robustness phase corresponding to the values of $\Lambda_1$ very close to zero. 
 
In Fig. \ref{fig:fig1}  \textbf{(B)}, we also compare the ADMFT's prediction of the steady-state ($m_\infty, \lambda, \mu$) with results from direct simulations of the original system ($m^{(s)}_\infty, \lambda^{(s)}, \mu^{(s)}$). We find a fairly good qualitative agreement between our theory and the numerics in terms of the critical transitions as well as the behaviours of $\lambda$ and $\mu$ (apart from small differences in their values from $ \lambda^{(s)}$ and  $\mu^{(s)}$ around the critical points). However, there is a large discrepancy in terms of the  averaged activity $m_\infty$ for $\sigma < \sigma_c^{(2)}\simeq 0.92$. We verified that this discrepancy is neither due to a finite-time integration of the dynamics (as well as finite time-separation) nor a finite-size effect as we have simulated for $N$ up to $N=1500$, but the results remain essentially similar to those of smaller $N=300, 600, 900$.
This discrepancy hence  must result from the  assumptions underlying our ADMFT derivation. Specifically, while the couplings are binary in the original dynamics of Eq. \eref{general1},  they are  assumed to be drawn from Gaussian distributions  in the effective dynamics of Eq. \eref{closed1}. 
 
 \begin{table}
\centering
\caption{The model steady-state phases and their features}
\begin{tabular}{lrrrr}
\\
Phases &  $m_\infty$ & $\lambda$ & $\mu$ \\
robust &  $m_\infty >0$ & $\lambda>0$ & $\mu>0$ \\
non-robust, & $m_\infty >0$  & $\lambda=0$  & $\mu>0$ \\
para-attractor (PA) & $m_\infty =0$ & $\lambda=0$ & $\mu=0$

\end{tabular}
\label{tabel:zscore}
\end{table} 

Figures   \ref{fig:fig1}  \textbf{(C)}--\textbf{(F)} show detailed behaviours of the order parameters  as functions of $T_J = 1/\beta$ and $\sigma$.  
 Both $m_\infty$ and $\mu$  only undergo an ordered/disordered transition as $\sigma$ is increased beyond $\sigma_c^{(2)}(T_J)$ at  fixed $T_J=\beta^{-1}$. Below some  $T^{(\lambda)}_J$, the robust  phase with $\lambda=1$ emerges within an intermediate range  $\sigma \in \big[\sigma_c^{(1)}(T_J), \sigma_c^{(2)}(T_J)\big]$. Importantly,  $\sigma_c^{(1)}(T_J)>0$.   These results are in agreement with  the model \cite{KanekoPloSOne2007}, where the robustness to noise  that evolved at intermediate noise is observed to be lost with decreasing noise. While the detailed mechanism for such a loss of robustness is beyond the scope of this paper, numerical results reported in \cite{KanekoPloSOne2007} show that this phenomenon appears as a consequence of a broad distribution of fixed points of  the original adaptive dynamics in Eq. \eref{general1} which exist at low noise. This suggests that the achievement of robustness at $\sigma_c^{(1)}(T_J)$ might be related to the so-called ``order-by-disorder phenomena'' known to occur in geometrically frustrated systems such as spin-glasses \cite{Villain}, where stochastic noise can result in  a noise-induced spontaneous symmetry breaking for the set of \emph{degenerated} orbits \cite{Hanai}. If noise is increased beyond $\sigma_c^{(2)}(T_J)$,    a transition from the robust to para-attractor phase occurs due to the dominant effect of noise. In the para-attractor region, both phenotypic and genotypic values become zero, indicating the unique state is neither robust nor functional. 
\begin{center}
 \begin{figure}[t]
 \centering
\includegraphics[scale=0.25]{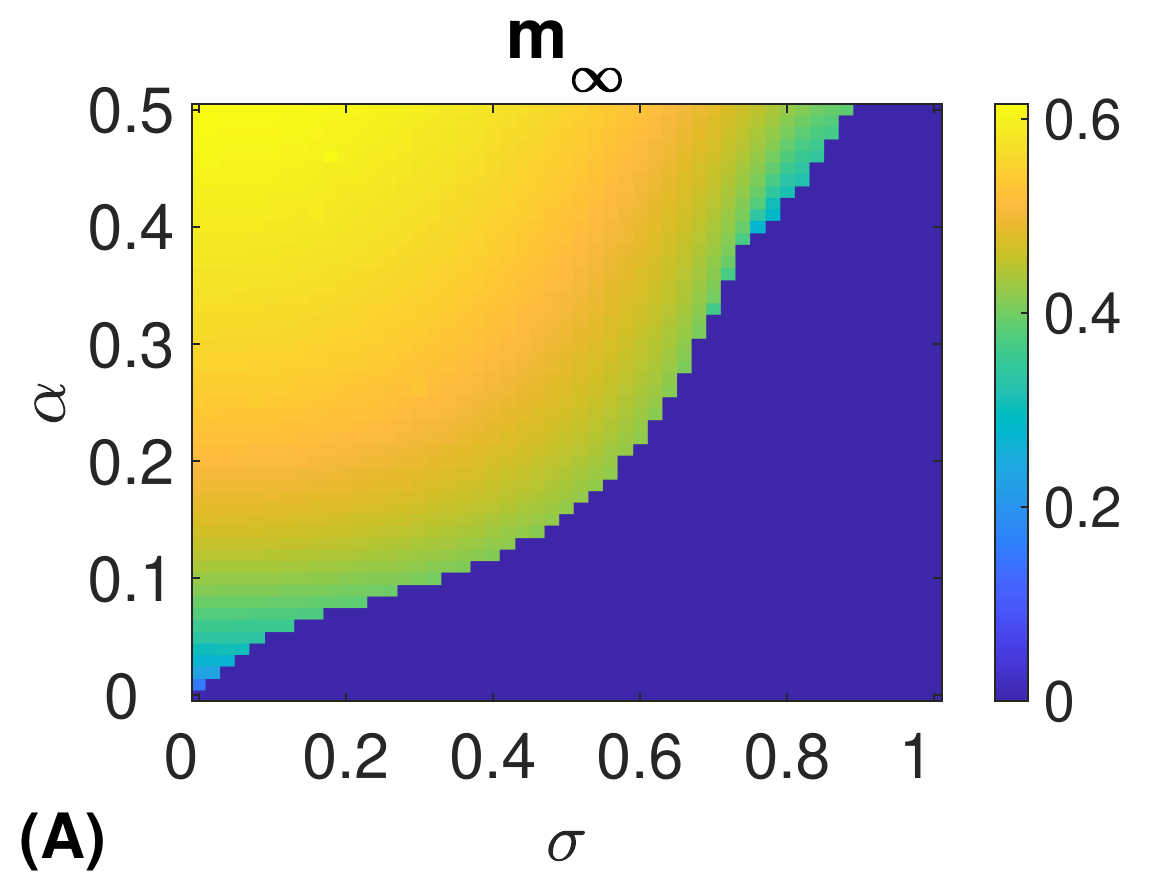}
\includegraphics[scale=0.25]{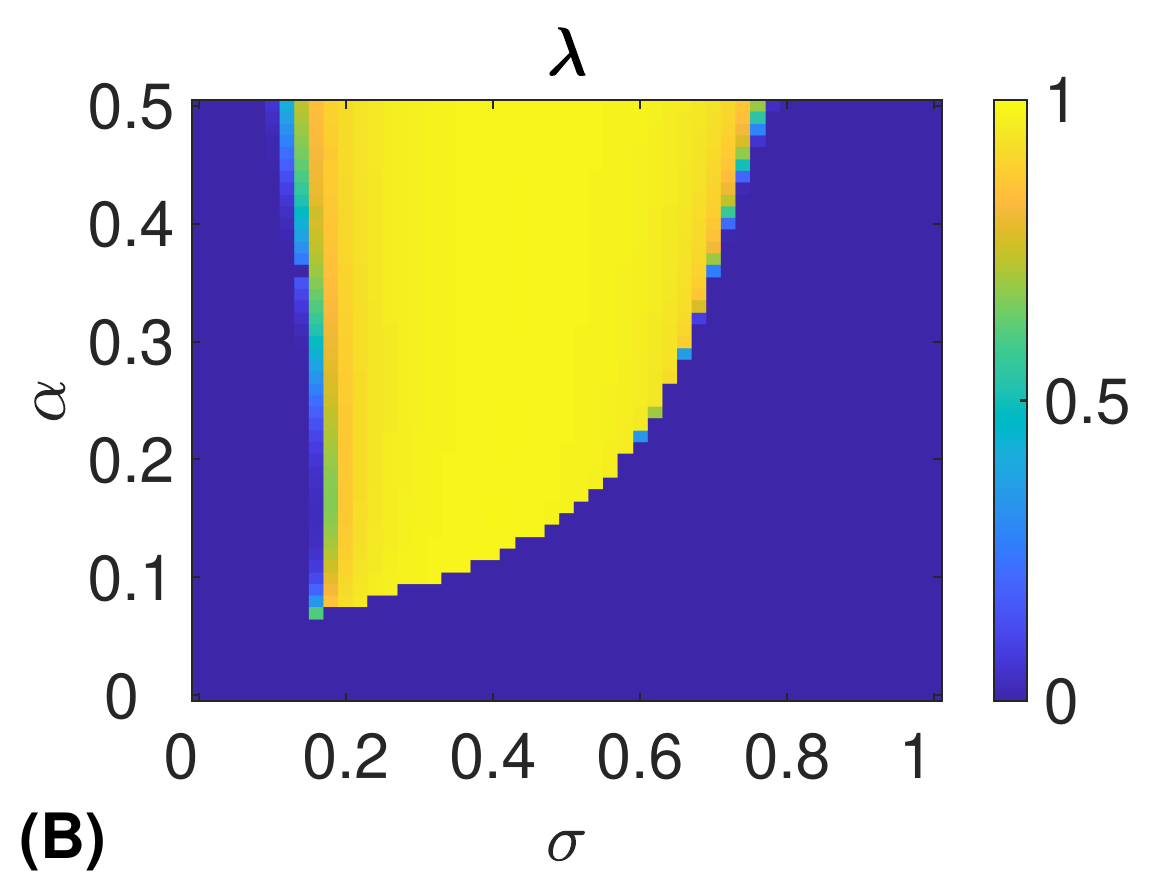}

\includegraphics[scale=0.25]{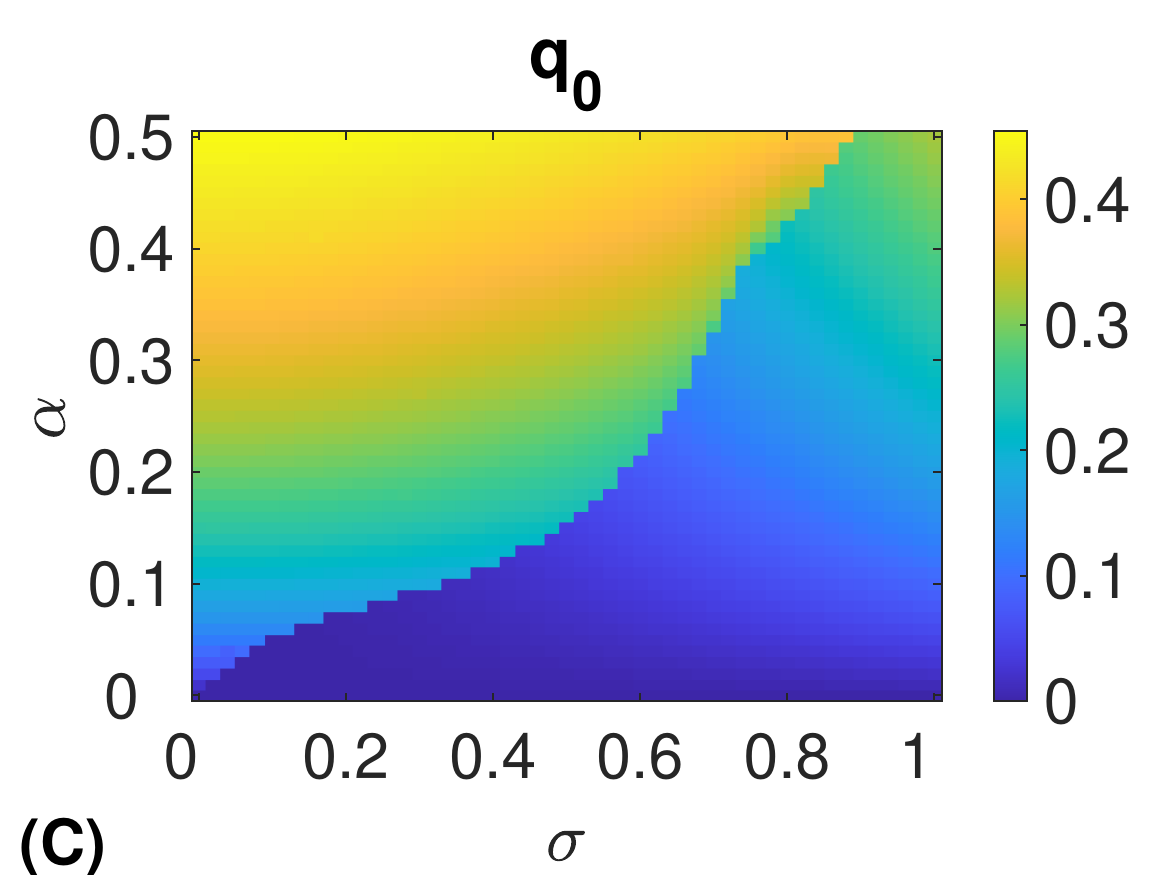}
\includegraphics[scale=0.25]{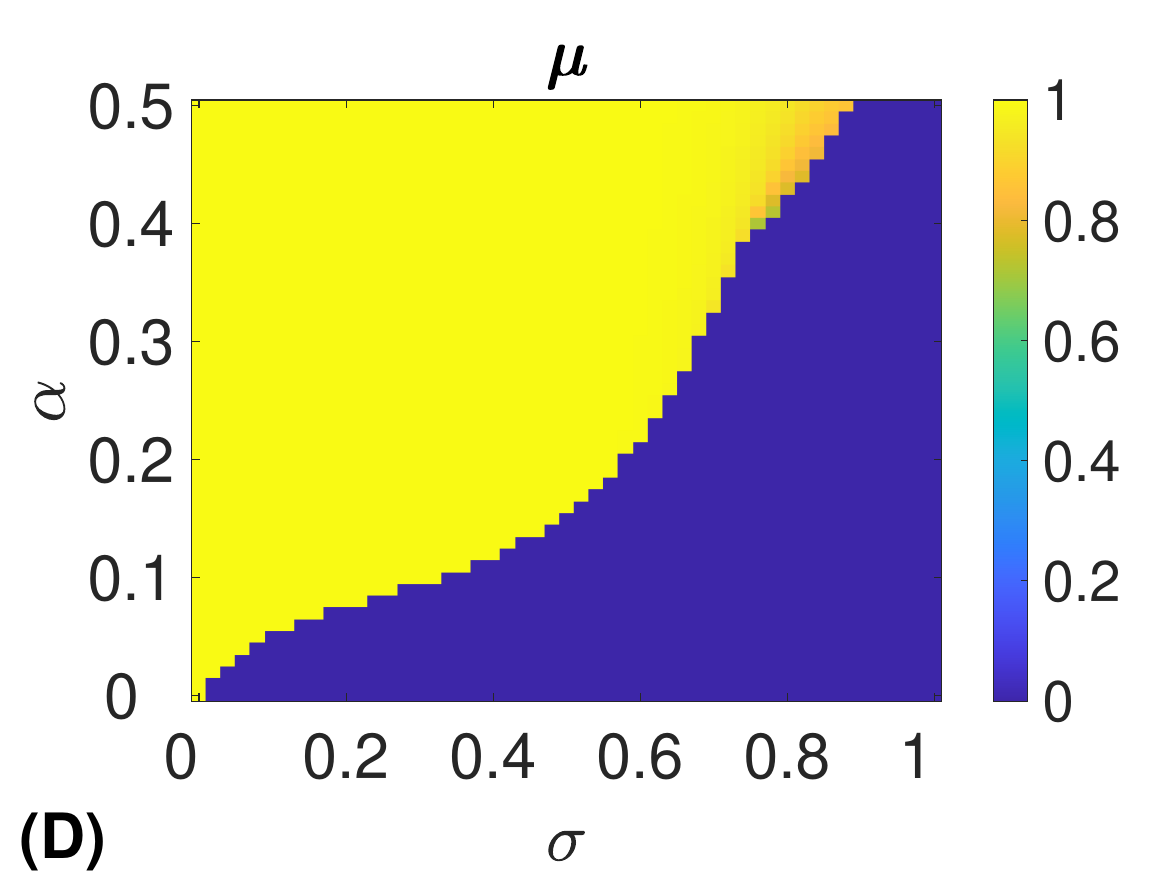}
\includegraphics[scale=0.25]{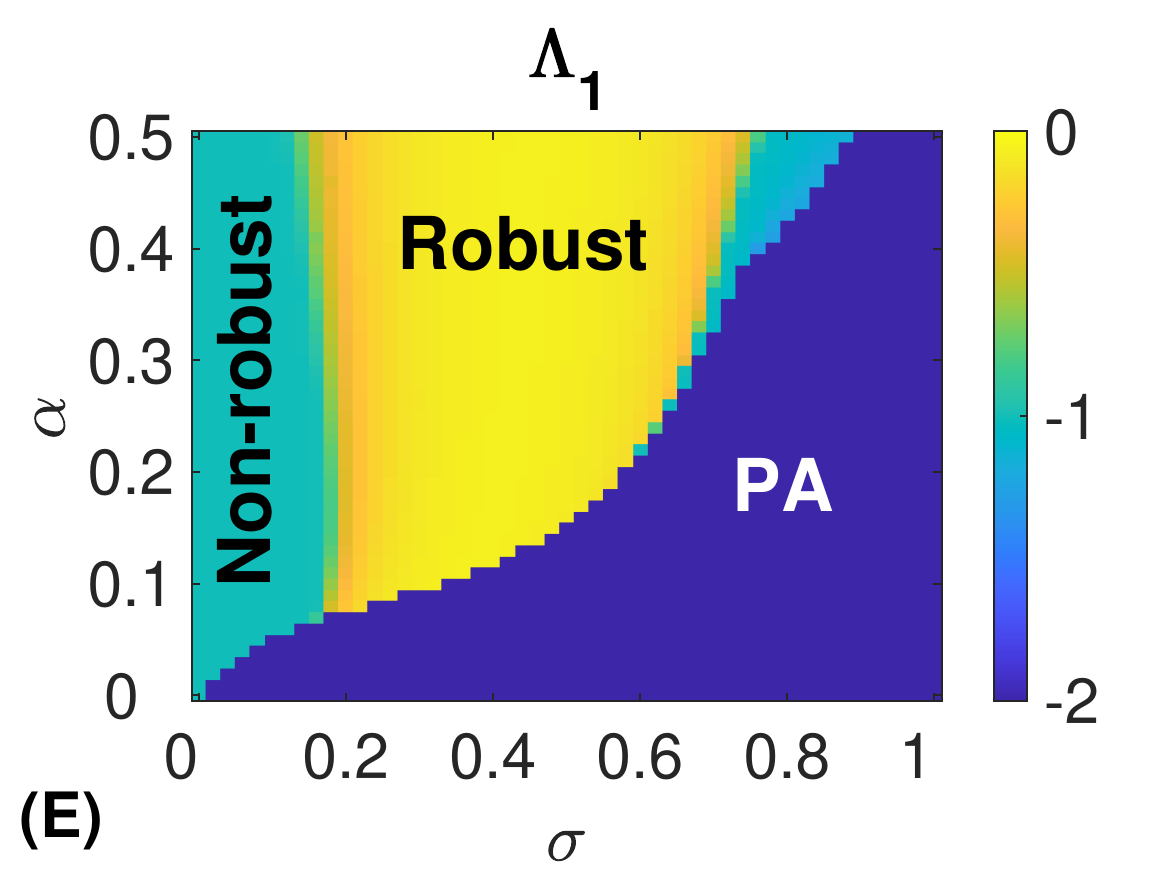}
\caption{Order parameters as function of $\alpha = N_o/N_t$ and $\sigma$  at $\nu=1$ and $\beta = 10$. \textbf{(A)}    Averaged activity of target genes $m_\infty$.  \textbf{(B)}  The average of target vs non-target coupling $\lambda$. \textbf{(C)} The intrinsic variance in the activity of target genes $q_0$.   \textbf{(D)} The average of target vs target coupling $\mu$. \textbf{(E)} $\Lambda_1$ as function of $\alpha = N_o/N_t$ and $\sigma$. The value of $\Lambda_1$ is used to distinguish the robust, non-robust,
and para-attractor phases.
} 
\label{fig:fig3}
\end{figure}
\end{center}

Next, in \Fref{fig:fig3}, we show that the existence of these three phases as well as the signatures of this phase transition towards robustness carry over to other $\alpha\neq 0.5$, but with the width of the robust region depending on $\alpha$.  We hence plot the order parameters' phase diagrams in terms of  $\alpha$ and  $\sigma$ in \Fref{fig:fig3} \textbf{(A)}-\textbf{(D)} and then use $\Lambda_1$ in  \Fref{fig:fig3} \textbf{(E)} to summarise  the system distinct phases.  At a given noise level $\sigma$, the system reaches a state with non-zero fitness $m_\infty>0$ for $\alpha>\alpha_c(\sigma)$. Note that $\alpha_c(\sigma)$ grows with $\sigma$, indicating the necessity of having a sufficient fraction of non-target genes to achieve high-fitness values at large noise. This enhancement of robustness with the help of non-target genes is also observed in the behavior of the genotypic variable $\lambda$, where  the robust region with $\lambda > 0$ only exists above $\alpha \simeq 0.07$ and
broadens with increasing $\alpha$.  Note that due to our choice of fully-connected networks, the fixed-point assumption only holds for $\alpha\leq 0.5$, as can be shown via a local stability analysis given in Appendix E. We  expect that in realistic genetic networks that are \emph{sparse}, such a restriction are not necessary. Therefore, one can consider $\alpha > 0.5$ and even $\alpha\gg 1$ in that case.

\section{Discussion}
 In this paper, we constructed
the ADMFT framework that is applicable to a wide range of  adaptive systems, in which slow adaptation of one type of degrees of freedom occurs in response to fast changes in the state of the other. We demonstrated our approach within  the context of genotype-phenotype evolution, where we found a  transition from  robust- to nonrobust phase with  decreasing noise in networks of fully-symmetric intergroup couplings. This happens due to a trade-off between phenotype and genotype that leads to a strong non-monotonic behavior of the genotypic value $\lambda$.   A comprehensive picture of the model behavior for any symmetry level $\nu\in[-1,1]$ with possible limit-cycle and chaotic attractors will be addressed in future work. In the meanwhile, we can speculate that, as the intergroup interactions become more non-reciprocal, the system is expected to achieve a lower averaged activity than in the case of reciprocal interactions, thus reducing such non-monotonicity.  A robust phase hence might exist at low noise.  In the fully symmetric case,  we also found the emergence of an outlier eigenvalue from  the genotype-phenotype feedback. Such  eigenmode controls the long-time behavior of the gene-expression dynamics, which is consistent with the dimensional reduction of phenotypic dynamics widely observed in studies based on genetic algorithms  \cite{Furusawa2018, Sato2020, Sato2023}. 
  
For a given environment, our work  confirms the loss of robustness 
 at low noise and the beneficial role of noise in the evolution of robustness as suggested in \cite{KanekoPloSOne2007}.   In particular,  in the robust phase where  $m_\infty>0$,  a proportionality between the response to environmental stochasticity   and the response to mutation 
  can be expected to arise as a consequence of the Hebbian-learning in Eq. \eref{h_for_Jtt_relative}. Such a proportionality might lead to a correlation between phenotypic changes due to genetic variation  and those in response to environmental perturbations, as discussed in \cite{KanekoPloSOne2007,Sato2003,Kaneko2006,Ciliberti, Sakata2020, Pham2022, Tang2021, Landry, Silva-Rocha, Uchida}. 
 It would be interesting to extend our approach to adaptation in fluctuating environments, where the environment alternates between different optimal  phenotypes. 

In comparison to other recent extensions of DMFT   for neural dynamics that focus on local learning rules, such as activity-dependent plasticity \cite{clark2023theory} and  pattern-based learning    \cite{Pereira-Obilinovic}, our approach derives the adaptation rules for the coupling matrix $\mathbf{J}$ from  a global fitness function. Nevertheless, 
 persistent fluctuations 
 are observed in the robust phase of the presented GRN model are similar to chaos with retrieval in \cite{Pereira-Obilinovic}.
In this regard, the ADMFT suggests the relevance of noise to shape robust memory by maintaining  a finite overlap with the stored patterns. 
Future work using ADMFT might shed light on the role of noise in other dynamics with timescale separation, such as Pavlov learning \cite{Agliari_pavlov} or learning dynamics of Restricted Boltzmann Machines \cite{fachechi2024}. 
 
\bigskip

\ack
We acknowledge support from Novo Nordisk Foundation (0065542) and would like to thank the anonymous reviewers, Albert Alonso, David Saad, Edo Kussell, Fernando  Metz,  Richardo Rao, Tarek Tohme, Valentina Ros and Yuhai Tu for helpful comments.

\appendix
\section*{Appendices}
\subsection*{A. Derivations of the effective couplings among the target units}
In the following we consider 2 groups of units: target ($i \in \mathcal{T}$) and non-target ($ i\in \mathcal{O}$).  From now on  by $\mathbf{x}$ we mean only $x_{i}$, for $i \in \mathcal{T}$. To avoid confusion, the state vector of all non-target units $x_{i}$, for $ i\in \mathcal{O}$, is denoted by  $\mathbf{y}$. Due to this partition, there are 4 different types of interactions, namely,   $J_{ij}^{(tt)}$ for $i \in \mathcal{T}$ and $j \in \mathcal{T}$;   $J_{ij}^{(oo)}$ for $i  \in \mathcal{O}$ and  $j \in \mathcal{O}$;  $J_{ij}^{(to)}$ for $i \in \mathcal{T}$ and $j \in \mathcal{O}$; $J_{ij}^{(ot)}$ for $i \in \mathcal{O}$ and $j \in \mathcal{T}$.
We have 
\numparts
\begin{eqnarray}
 \frac{ \partial}{\partial t}x_k &=& - x_k+ F\Big(\sum_{j=1, j\neq k}^{N_t} J^{(tt)}_{kj}(\tau) x_j+ \sum_{\ell=1}^{N_o} J^{(to)}_{k\ell}(\tau) y_\ell\Big) + \xi_k\,. 
\label{target_subset} \\
 \frac{ \partial}{\partial t}y_k &=& - y_k + F\Big(\sum_{j=1, j\neq k}^{N_o} J^{(oo)}_{kj}(\tau) y_j+ \sum_{\ell=1}^{N_t} J^{(ot)}_{k\ell}(\tau) x_\ell \Big) + \xi_k\,,
\label{non_target_subset}
\end{eqnarray}
\endnumparts
The presence of a non-target unit $\ell \in \mathcal{O}$ with its dynamics modifies the dynamic of those target units $k\in \mathcal{T}$ connected to it. Therefore, to derive a closed set of  dynamical equations for the subset of target variables exclusively,    we consider the following \emph{approximate} dynamics for the  target units whose  effective interactions are $J^{(e)}_{kj}(\tau)$ for $(k,j)\in \mathcal{T}$: 
\begin{equation}
      \frac{ \partial}{\partial t}x_k(t,\tau) = - x_k(t,\tau) + F\Big(\sum_{j=1, j\neq k}^{N_t} J^{(e)}_{kj}(\tau) x_j(t,\tau) \Big) + \xi_k(t,\tau)\,\quad     \label{target_dynamics}
\end{equation}
and further assume that  steady state solutions of Eq. \eref{target_dynamics} is the same as those of the original dynamics Eq. \eref{target_subset}. Below, we  provide a condition under which this assumption should be valid. At first, we introduce the following ansatz for $J^{(e)}_{kj}(\tau)$ 
\begin{equation}
    J^{(e)}_{kj}(\tau) = J^{(tt)}_{kj}(\tau) + \Delta J_{kj}(\tau) \,,\qquad k,j\in \mathcal{T} \label{signal_to_noise}
\end{equation}
where, apart from their direct interaction $J^{(tt)}_{kj}(\tau)$ ($J^{(tt)}_{kj}(\tau)\in \{-1,1\})$,   $\Delta J_{kj}(\tau)$ is the part of the effective coupling that is induced by all their common \emph{non-target} neighbors $\ell \in \mathcal{O}$ via $J^{(to)}_{k\ell}(\tau)$ and $J^{(ot)}_{\ell j}(\tau)$. 
As
at generation $\tau$,  both  $\mathbf{x}$ and $\mathbf{y}$   reach their corresponding attractors   defined component-wise from Eqs. \eref{target_subset}-\eref{non_target_subset} as
\begin{equation} x_k(\tau) := \lim_{t\rightarrow\infty} x_k(t,\tau) 
= F\Big(\sum_{j\in \mathcal{T}} J_{kj}^{(tt)}(\tau) x_j(\tau) + \sum_{\ell \in \mathcal{O}} J^{(to)}_{k\ell}(\tau)y_\ell(\tau)\Big) 
\label{attractor_definition}
\end{equation}
$$y_\ell(\tau) = F\Big(\zeta_\ell^{(o)} + \zeta_\ell^{(t)} \Big)\,,\quad \zeta_\ell^{(o)} \equiv \sum_{\ell'\in \mathcal{O}} J^{(oo)}_{ \ell \ell'}(\tau)y_{\ell'}(\tau)\,,\quad\zeta_\ell^{(t)} \equiv \sum_{j\in \mathcal{T}} J_{ \ell j}^{(ot)}(\tau)x_j(\tau)$$
where $\zeta_\ell^{(o)}$ and $\zeta_\ell^{(t)}$ are the contributions of $\ell'\in \mathcal{O}$ and  $j\in \mathcal{T}$, respectively, to $y_\ell(\tau)$.
 If there is no feedback of phenotype on the couplings among the non-target units, then $h^{(oo)}_{\ell \ell'}(\tau) =0$, $\forall \tau$, and as a result, $J^{(oo)}_{\ell \ell'} $ remains random over generations. Therefore, we have $\zeta_\ell^{(o)} \ll \zeta_\ell^{(t)} \ll 1$, and 
$y_\ell(\tau) \simeq \zeta_\ell^{(t)} $.
 Substituting this $y_\ell(\tau)$ into Eq. \eref{attractor_definition}, we get 
the condition for the asymptotic attractor of Eq. \eref{target_dynamics} equals to that of  Eq. \eref{target_subset} 
\begin{equation} 
 x_k(\tau)  \simeq F\big( \sum_{j\in \mathcal{T}} \big[J^{(tt)}_{kj}(\tau) + \Delta J_{kj}(\tau)\big] x_j(\tau) \big)
 \label{approximate_attractor_definition}
 \end{equation}
  if and only if
\begin{equation}
     \Delta J_{kj}(\tau) =\sum_{\ell \in \mathcal{O}} J_{k\ell}^{(to)} (\tau)J_{\ell j}^{(ot)}(\tau)\,.
\label{relation}
 \end{equation}
\subsection*{B. Derivations of \Eref{h_for_Jto} and \Eref{steady_state_jto}}
From Eq. \eref{squared_fitness} we have the total change of the fitness if all the couplings but $J_{k\ell}^{(to)}(\tau)$ are fixed, i.e. $J_{i\ell}^{(to)}(\tau+1) = J_{i\ell}^{(to)}(\tau) $, $\forall\, i\neq k, i \in \mathcal{T}$:
\begin{equation}
\Delta\Psi = \Psi(J_{k\ell}^{(to)}(\tau+1) ) - \Psi(J_{k\ell}^{(to)}(\tau) ) = N_t^{-2}\sum_{i: i\neq k} \Big([x_kx_i]_{\tau+1} - [x_kx_i]_{\tau}\Big)
\end{equation}
Assuming  the distribution remains unchanged, i.e. $P_{\mathbf{J}(\tau+1)}(x_{\mathcal{T}}) \simeq P_{\mathbf{J}(\tau)}(x_{\mathcal{T}}) $, we arrive at the following, to the first order in $\Delta J^{(to)}_{k\ell}$, for $\Delta J^{(to)}_{k\ell} := \big[J^{(to)}_{k\ell}(\tau+1) - J^{(to)}_{k\ell}(\tau)\big] \rightarrow 0$, 
\numparts
\begin{eqnarray}
\Delta\Psi & \approx &  N_t^{-2}\sum_{i: i\neq k} \int d^{N_t}xP_{\mathbf{J}(\tau)}(x_{\mathcal{T}}) x_i\big[x_k(\tau+1) - x_k(\tau)\big] \\ &\simeq & N_t^{-2}\sum_{i: i\neq k} \int d^{N_t}xP_{\mathbf{J}(\tau)}(x_{\mathcal{T}}) x_i \left( \frac{\partial x_k}{\partial J^{(to)}_{k\ell}}\, \Delta J^{(to)}_{k\ell}\right)
\label{total_fitness_change}
\end{eqnarray}
\endnumparts
Using Eq. \eref{approximate_attractor_definition} and \eref{relation} for $F={\rm tanh}(\cdot)$ we have 
\begin{equation}
\frac{\partial x_k}{\partial J^{(to)}_{k\ell}} =  \big(1-x_k^2(\tau)\big)\sum_{j\in\mathcal{T}}J^{(ot)}_{\ell j}x_j(\tau)
\label{fitness_change}
\end{equation}
 Plugging  Eqs. \eref{fitness_change}   into Eq. \eref{total_fitness_change}, we obtain
\begin{equation}  \Delta \Psi = \Delta J^{(to)}_{k\ell} \big(1-x_k^2(\tau)\big)  N_t^{-2}\int d^{N_t}xP_{\mathbf{J}(\tau)}(x_{\mathcal{T}})  \sum_{i,j\in\mathcal{T}}J^{(ot)}_{\ell j}x_i x_j\label{h_for_Jto2}
  \end{equation}
If for $k\in \mathcal{T}, l\in \mathcal{O}$, we introduce
$$h_{k\ell}^{(to)}(\tau) := \big(1-x_k^2(\tau)\big)  N_t^{-2}\int d^{N_t}xP_{\mathbf{J}(\tau)}(x_{\mathcal{T}})  \sum_{i,j\in\mathcal{T}}J^{(to)}_{i\ell}(\tau) J^{(ot)}_{\ell j}(\tau)x_i x_j$$
then by noting that $\Delta J^{(to)}_{k\ell}= \sum_{i=1}^{N_t}\big[J^{(to)}_{i\ell}(\tau+1) - J^{(to)}_{i\ell}(\tau)\big]$, we get
\begin{equation}
 h_{k\ell}^{(to)}(\tau+1) - h_{k\ell}^{(to)}(\tau)  =\frac{\partial \Psi(J_{k\ell}^{(to)}(\tau) )}{\partial J_{k\ell}^{(to)}(\tau) }
 \label{fitness_gradient2}
\end{equation}
Substituting the form of $h_{k\ell}^{(to)}(\tau)$ from Eq. \eref{h_for_Jto_relative} into Eq. \eref{jto}  and then taking the limit $t,\tau\rightarrow \infty$, we arrive at Eq. \eref{steady_state_jto}.

 \subsection*{C. Derivation of the effective dynamics in \Eref{closed1}} 

 We use  an integral representation of the probability $P(\xi)$ of the mean zero white noise $\xi(t)$:
 $$P(\xi) \sim \exp\left[-\displaystyle\frac{1}{2} \int dt  dt'  \xi(t)\big[C_{\xi}(t,t')\big]^{-1} \xi(t') \right]\,,\qquad C_{\xi}(t,t') =\langle \xi(t)\xi(t')\rangle =  \sigma^2\delta(t-t')$$
to rewrite  any SDE of the form $\partial_t x= \mathcal{F}(t) + \xi(t)$ as [$i$ is the imaginary unit $i^2=-1$]
\begin{equation}
1= \int Dx D\hat{x} \exp\left[-\displaystyle\frac{1}{2} \int dt dt'\,  \hat{x}(t)C_{\xi}(t,t')\hat{x}(t')+  i \int dt\,  \hat{x}(t)\big[\partial_t x  - \mathcal{F}(t)   \big] \right]
\label{identity_for_Gaussian_noise}
\end{equation}
Let $\langle \cdot \rangle$ denote the average taken wrt the measure $\mathbb{P}\big(\{\mathbf{x}\}_{0,T_{\rm max}}\big)$ -- the distribution over an ensemble of trajectories of $\mathbf{x}(t;\tau)$ for $t\in [0, \infty)$ over $T_{\rm max}$ generations:
$$\mathbb{P}\big(\{\mathbf{x}\}_{0,T_{\rm max}}\big) := \mathbb{P}\Big[\big\{\mathbf{x}([0, t_f],\tau = 0)\big\}, \big\{\mathbf{x}([0, t_f],\tau = 1)\big\},\cdots, \big\{\mathbf{x}([0,t_f], \tau = T_{\rm max})\big\}\Big]_{t_f\rightarrow \infty}\,.$$
Since  at each generation $\tau \in\{0,1,\cdots,T_{\rm max}\}$, 
$\mathbf{x}(t,\tau)$ obey the first of Eq. \eref{general1},  we can   represent $ \mathbb{P}\big(\{\mathbf{x}\}_{0,T_{\rm max}}\big)$  by rewriting the  dynamics of $\mathbf{x}$
 in the presence of an external field $\bm{\theta}(t,\tau)$ with the help of Eq. \eref{identity_for_Gaussian_noise} as an identity:
\numparts \begin{eqnarray}
1= \int D\big[x\hat{x}f\hat{f}\big] \exp\Big\{i\sum_{\tau=0}^{T_{\rm max}} \sum_{k=1}^{N} \int dt \Big[S^{(0)}_k -  I_k\Big]\Big\} 
\label{identity}\\
S^{(0)}_k[x,\hat{x}, f,\hat{f}] = \hat{x}_k \Big[(\partial_t +1)x_k  - F\big(f_k\big) \Big] + \hat{f}_k\big[f_k-\theta_k\big] + i\sigma^2 \hat{x}^2_k/2   \label{S_k}\\   I_k[x,\hat{f}]=\sum_{j=1}^{N} J_{kj}(\tau) \hat{f}_k(t,\tau)  x_j(t,\tau) \label{I_k}
\end{eqnarray} \endnumparts
 where $D[x\hat{x}f\hat{f}] := \prod_{n=0}^{t_f/\Delta t}  \prod_{k=1}^{N}  \prod_{\tau=0}^{T_{\rm max}} Dx_k(n\Delta t, \tau)  D\hat{x}_k(n\Delta t, \tau) Df_k(n\Delta t, \tau)  D\hat{f}_k(n\Delta t, \tau)$, for $\Delta t\rightarrow 0$, denotes  the functional measure over all possible paths. 
The moment generating functional of $\mathbb{P}\big(\{\mathbf{x}\}_{0,T_{\rm max}}\big)$ with $\{\bm{\psi}(t,\tau)\}$ for $\tau = 0,\cdots, T_{\rm max}$ is: 
\begin{equation}
Z[\bm{\psi}] = \left\langle \exp\left\{i\sum_{\tau=0}^{T_{\rm max}} \sum_{k=1}^{N} \int dt\, \psi_{k}(t,\tau)x_k(t,\tau) \right\} \right\rangle
\label{moment_generating_function}
\end{equation}
When  the distribution of noise is $p(\tilde{\xi}) = \big[1- {\rm tanh}^2(\tilde{\xi})\big]/2$, we can rewrite the second dynamics in
 Eq. \eref{general1} as a   master equation  for the distribution $\tilde{P}\big(\mathbf{J}(\tau)\big)$ of  configurations $\mathbf{J}(\tau)$ \cite{COOLEN2001}:
 \begin{equation}
     \tilde{P}(\mathbf{J}(\tau+1)) = \sum_{\{\mathbf{J}(\tau)\}}\tilde{P}(\mathbf{J}(\tau))\prod_{k\neq j}\frac{e^{\displaystyle\beta h_{kj}(\tau) J_{kj}(\tau+1)}}{2{\rm cosh}[\beta h_{kj}(\tau)]}   \label{master_equation}
 \end{equation}
This means the actual choice of value of $J_{kj}(\tau+1)$ is probabilistic and drawn from a distribution over two possibilities, namely,  
$J_{kj}(\tau+1) = 1$ and  $J_{kj}(\tau+1) = -1$,  with probability  
$\propto \exp\big\{\beta h_{kj}(\tau) J_{kj}(\tau+1)\big\}$.
Without loss of generality, we take  $\tilde{P}(\mathbf{J}(0))$ as a uniform distribution. In order  to separate the feedback fields at any generation, let us introduce  $\tilde{h}_{kj}(\tau)$ as the value that $h_{kj}(\tau)$ admits at the $\tau$-th generation. This can be done by inserting the following identity to Eq. \eref{master_equation}:
\begin{equation}
    1= \int D\big[h\hat{h}\big]\exp\Big\{i\sum_{\tau=0}^{T_{\rm max}-1} \sum_{k,j} \hat{h}_{kj}(\tau)\big[h_{kj}(\tau)- \tilde{h}_{kj}(\tau)\big]\Big\} \label{identity_feedback_fields}
\end{equation}
For example, in the case of the genotype-phenotype  model in Section 4,  $\tilde{h}_{kj}(\tau)$ is given by \Eref{h_for_Jtt} if $(k,j) \in \mathcal{T}$ and \Eref{h_for_Jto} if $k\in \mathcal{T}$ and $j\in \mathcal{O}$. As  seen from that example,  in general, $\tilde{h}_{kj}(\tau)$ depends on the joint distribution of the fields $\mathbf{x}$ and $\mathbf{J}$. 

Let 
$\mathbb{E}[\cdot]$ denote the ensemble average wrt the joint distribution of trajectories in the combined space of $\mathbf{x}$ and $\mathbf{J}$. This measure can be obtained by inverse Fourier transform of the moment-generating functional $Z[\bm{\psi}, \bm{\Psi}]$, obtained by plugging Eqs. \eref{identity}-\eref{identity_feedback_fields}, altogether into Eq. \eref{moment_generating_function}:
\numparts 
\begin{eqnarray}
    Z[\bm{\psi}, \bm{\Psi}]  &=&  \int D\big[ x\hat{x} f\hat{f} h \hat{h} \big]   \sum_{\{\mathbf{J}(0)\}} \cdots \sum_{\{\mathbf{J}(T_{\rm max})\}} e^{\displaystyle\mathcal{L}} \label{moment_generating_function2}\\ \mathcal{L}&=&i\sum_{\tau=0}^{T_{\rm max}}\sum_{k=1}^{N} \int dt \Big[S^{(\psi)}_k -   I_k\Big] + \sum_{k\neq j}\big[I_{kj} + J_{kj}(0)\Psi_{kj}(0)\big]\label{moment_generating_function4}
 \end{eqnarray}
  \endnumparts
  where  
$$S^{(\psi)}_k = S^{(0)}_k + \psi_{k}(t,\tau)x_k(t,\tau)$$
$$L_{kj}(\tau) = i\hat{h}_{kj}(\tau)\big[h_{kj}(\tau)- \tilde{h}_{kj}(\tau)\big]- \ln \big(2{\rm cosh}\, [\beta h_{kj}(\tau)]\big)$$
$$I_{kj}  = \sum_{\tau=0}^{T_{\rm max}-1}  \Big\{ L_{kj}(\tau) + J_{kj}(\tau+1)\big[\Psi_{kj}(\tau+1) +\beta h_{kj}(\tau)\big] \Big\} $$
Denoting the $J$-independent and $J$-dependent part of $\mathcal{L}$  by $\mathcal{L}_0$ and $\mathcal{L}_J:= \sum_{k\neq j}  \mathcal{L}_{kj}^{(J)} $, respectively, the exponential in $Z[\bm{\psi}, \bm{\Psi}]$ can be written as: 
\numparts 
\begin{eqnarray}
\mathcal{L}&=&\mathcal{L}_0 +  \mathcal{L}_J\label{generating_function3b}\\
 \mathcal{L}_0 &=&  i\sum_{\tau=0}^{T_{\rm max}} \sum_k \int dt S_k^{(\psi)}(t,\tau) + \sum_{\tau=0}^{T_{\rm max}-1} \sum_{k\neq j}L_{kj}(\tau)  \\
\nonumber\mathcal{L}_{kj}^{(J)}   =& + & J_{kj}(0)\Big[\Psi_{kj}(0)- i\int dt \hat{f}_k(t,0) x_j (t,0)\Big] \\ &+&  \sum_{\tau=1}^{T_{\rm max}} J_{kj}(\tau)\Big\{ \Psi_{kj}(\tau) + \beta h_{kj}(\tau-1)- i\int dt \hat{f}_k(t,\tau) x_j (t, \tau)\Big\} \qquad  \quad 
\end{eqnarray}
\endnumparts 

From now on, we shall  compute $Z[\bm{\psi}, \bm{\Psi}]$, for the above dynamics of Eq. \eref{target_dynamics}. This means that we need to    insert $J_{kj}(\tau) = J^{(e)}_{kj}(\tau)$ (with $ \Delta J_{kj}$  from Eq. \eref{relation}) into the expression of $I_k[x,\hat{f}]$  in \Eref{I_k}. As we consider  fully-connected networks, if $N_t\rightarrow \infty$,  the couplings $ J^{(e)}_{kj}$ should be rescaled by $1/\sqrt{N_t}$ to ensure a sensible thermodynamic limit.  However, due to a  low-rank structure that can emerge from the adaptation dynamics of $J^{(tt)}_{kj}$,  a proper scaling of $J^{(e)}_{kj}$  is $J^{(e)}_{ij}\rightarrow J^{(e)}_{ij}/N_t$. 
We define the order parameters:
\numparts 
 \begin{eqnarray}
 w_{kj}(\tau) &= &\frac{1}{N_t}\int dt  \hat{f}_k(t,\tau) x_j(t,\tau)
  \label{auxiliary1}
\\  m(t,\tau) & =&  \frac{1}{N_t}\sum_{j\in \mathcal{T}} x_j (t,\tau) 
  \label{auxiliary2}
  \\  g(t,\tau) &=& \frac{1}{N_t}\sum_{k\in \mathcal{T}}  \hat{f}_{k}  (t,\tau) 
  \label{auxiliary3} 
  \\ q(t,t',\tau) &= &  \frac{1}{N_t}\sum_{k\in \mathcal{T}}  x_k(t,\tau) x_k(t',\tau)
  \label{auxiliary4}
\\ Q(t,t',\tau) &= & \frac{1}{N_t}\sum_{k\in \mathcal{T}}  \hat{f}_k(t,\tau) \hat{f}_k(t',\tau) 
  \label{auxiliary5}
  \\ K(t,t',\tau) &= &  \frac{1}{N_t}\sum_{k\in \mathcal{T}}  x_k(t,\tau) \hat{f}_k(t',\tau)\,,
  \label{auxiliary6}
\end{eqnarray}
\endnumparts 
Let  $[a]_{t,\tau}$ denote  $a(t,\tau)$.  We remark that  
     $\sum_{k\neq j} w_{kj}(\tau) = N_t\int dt\,  m(t,\tau) g(t,\tau)$. Thus,
\begin{eqnarray*}
 1 &= \int D[m\hat{m} g\hat{g}\hat{\mu}w]  \exp\left\{ i \int dt\, \left[ \hat{\mu}\Big(\sum_{k\neq j\in \mathcal{T }} w_{kj} - mgN_t   \Big) \right]_{t, \tau} \right\} \\ &\times \exp\left\{ i \int dt\, \left[\hat{m}\Big(mN_t- \sum_{j\in \mathcal{T}}   x_j \Big) + \hat{g}\Big(gN_t - \sum_{k\in \mathcal{T}} \hat{f}_{k}   \Big) \right]_{t, \tau} \right\}\,.
   \end{eqnarray*}
Using this identity  to perform the sum over $\{\mathbf{J}(\tau)\}$ in Eq. \eref{moment_generating_function2}, we can write the moment generating functional $Z[ \bm{\psi},\mathbf{\Psi}]$  as follows:
\begin{eqnarray}
 Z[ \bm{\psi},\mathbf{\Psi}]& =&\int D\big[ x\hat{x} f\hat{f} h \hat{h} \big]\exp\Big\{\displaystyle\hat{\mathcal{L}}_0+ \sum_{\tau=0}^{T_{\rm max}-1} [B(\tau) +  D(\tau)]\Big\}\label{generating_function3a} \\
\hat{\mathcal{L}}_0&\equiv&   \mathcal{L}_0 +\sum_{k\neq j \in \mathcal{T}} \ln 2\,{\rm cosh}\big(\Psi_{kj}(0)-iw_{kj}(0)\big) \qquad \quad \end{eqnarray}
where $B(\tau)$ and $D(\tau)$ shall be obtained by summing over the respective parts 
$J_{kj}^{(tt)}(\tau)$ and $\Delta J_{kj}(\tau)$ of $J^{(e)}_{kj}$ given in \Eref{signal_to_noise}.
The sum over $\Big\{J_{kj}^{(tt)}(\tau)\Big\}$ is  straightforward to be carried out and yields
\begin{eqnarray*} \nonumber  B(\tau) &=   \ln \left\{ \int D[\hat{\mu}w m g] \prod_{k\neq j \in \mathcal{T}} 2{\rm cosh}\,\left(\frac{\Psi_{kj}(\tau+1) +\beta h^{(tt)}_{kj}(\tau) -i w_{kj}(\tau+1)}{N_t}  \right) \right\}\\ \nonumber&+ \ln\left\{  \exp\Big( i\hat{\mu}(\tau+1) \int dt \,\Big[ g \Big(mN_t  -\sum_{k\in \mathcal{T}} x_{k}  \Big)+ m\Big(gN_t-  \sum_{k\in \mathcal{T}}   \hat{f}_k \Big)\Big]_{t, \tau+1}       \Big)  \right\}
\end{eqnarray*}
To compute the sum over $\Big\{\Delta J_{kj}(\tau) \Big\}$, for $\Delta J_{kj}(\tau) = \sum_{ l \in \mathcal{O}}J_{kl}^{(to)}(\tau)J_{lj}^{(ot)}(\tau) $ we introduce:
\begin{eqnarray*}  \exp\left(\sum_{k\neq j\in \mathcal{T}} \tilde{I}_{kj}\right)&:= \exp\left\{\displaystyle-\frac{i}{N_t} \sum_{k\neq j\in \mathcal{T}} \sum_{ l\centernot \in \mathcal{T}}  \int dt \Big[\hat{f}_kx_jJ_{kl}^{(to)}(\tau)J_{lj}^{(ot)}(\tau)  \Big]_{t, \tau}   \right\} \\
 &=\exp\left\{-i \sum_{l\centernot \in \mathcal{T}} \int dt  \left[\sum_{k \in \mathcal{T}} \frac{J_{kl}^{(to)}(\tau)}{\sqrt{N_t}} \hat{f}_k  \right]_{t, \tau}    \left[\sum_{j \in \mathcal{T}} \frac{J_{lj}^{(ot)}(\tau) }{\sqrt{N_t}}x_j \right]_{t, \tau} \right\}\\ &=
\int \prod_{l\centernot \in \mathcal{T} } Dy_l Dz_l e^{  -i \int dt \,z_l(t, \tau)  y_l(t, \tau) }  \\ &\times\int dt  \left[ \delta\Big(z_l- \sum_{k\in \mathcal{T}} \frac{J_{kl}^{(to)}\hat{f}_k}{\sqrt{N_t}} \Big)  \delta\Big(y_l- \sum_{j\in \mathcal{T}} \frac{J_{lj}^{(ot)} x_j}{\sqrt{N_t}} \Big) \right]_{t, \tau} 
\end{eqnarray*}
The delta functions in this expression can be represented using the conjugate fields $\hat{y}_l$ and $\hat{z}_l$ as follows
$$ \int D\big[y\hat{y}z\hat{z}\big]  \exp\left\{\displaystyle i \sum_{l\centernot \in \mathcal{T}} \int dt[N_t^{1/2}(z_l\hat{z}_l + y_l\hat{y}_l)-z_ly_l ]_{t,\tau} -i \sum_{k\in \mathcal{T},l\centernot \in \mathcal{T}} \int dt \,  \big[   J_{kl}^{(to)}\hat{f}_k \hat{z}_l +  J_{lk}^{(ot)}x_k\hat{y}_l\big]_{t,\tau} \right\} $$
Substituting $\hat{y}_l = z_l/\sqrt{N_t}$ and $\hat{z}_l = y_l/\sqrt{N_t}$ that are obtained by varying the exponent wrt $y_l(t,\tau)$ and $z_l(t,\tau)$, respectively and denoting
$$X_{kl}(\tau) \equiv N_t^{-1/2}\Big\{  J_{kl}^{(to)}(\tau)\big[\Psi_{kl}(\tau)+ \beta h_{kl}^{(to)}(\tau-1)\big]+J_{lk}^{(ot)}(\tau)\big[\Psi_{lk}(\tau)+ \beta h_{lk}^{(ot)}(\tau-1)\big]\Big\} \,,$$
we have 
\numparts 
\begin{eqnarray}
   D(\tau) &= \ln\left\{\sum_{ \big\{J^{(to)}_{kl}(\tau),  J^{(ot)}_{lk}(\tau) \big\}}    \exp\Big(\sum_{ k \in \mathcal{T}, l\centernot \in \mathcal{T} } X_{kl}(\tau)  +\sum_{k\neq j\in \mathcal{T}} \tilde{I}_{kj} \Big) \right\} 
 \\ &= \ln\left\{\int D\big[yz q\hat{q}Q\hat{Q}K\hat{K}\big]\, e^{\displaystyle N_t\big[     i\Phi +   \Omega \big]} \exp\Big( \sum_{l\centernot \in \mathcal{T}} \int dt \,z_l(t,\tau) y_l(t,\tau)\,\Big) \right\} \qquad \qquad 
\end{eqnarray} 
\endnumparts 
where we  introduced 
\begin{eqnarray} \nonumber
\Phi = & -\frac{1}{N_t}  \sum_{k\in \mathcal{T}} \int dt dt'\, \hat{q}(t,t',\tau)    x_k(t, \tau)x_k(t', \tau)  \\\nonumber
&  -\frac{1}{N_t}  \sum_{k\in \mathcal{T}} \int dt dt'\,\left[\hat{Q}(t,t',\tau) \hat{f}_k(t, \tau)\hat{f}_k(t', \tau)+ \hat{K}(t,t',\tau)  \hat{f}_k(t', \tau)x_k(t, \tau) \right]\\ \nonumber&+  \int dt dt'\left\{ q(t,t',\tau)\left[\hat{q}(t,t',\tau) + \frac{i }{2N_t} \,\sum_{l\centernot \in \mathcal{T}} z_l(t,\tau)z_l(t',\tau) \right] \right\} \\ \nonumber &+ \int dt dt'\left\{ Q(t,t',\tau)\left[\hat{Q}(t,t',\tau)  +  \frac{i}{2N_t} \,\sum_{l\centernot \in \mathcal{T}} y_l(t,\tau)y_l(t',\tau)\right]\right\} \\  &+ \int dt dt'\left\{ K(t,t',\tau)\left[ \hat{K}(t,t',\tau) + \nu\frac{i}{N_t} \,\sum_{l\centernot \in \mathcal{T}}z_l(t,\tau)y_l(t',\tau) \right] \right\} 
\end{eqnarray}

\begin{eqnarray} \nonumber
\Omega  &= \frac{1}{2N_t^2} \sum_{k \in \mathcal{T},l \centernot \in \mathcal{T}}   \Big(\Psi_{kl}(\tau+1) + \beta h^{(to)}_{kl}(\tau)+\Psi_{lk}(\tau+1) + \beta h^{(ot)}_{lk}(\tau)\Big)^2  \\  &- \frac{i}{N_t^2} \sum_{k \in \mathcal{T},l \centernot \in \mathcal{T}} \Big[\Psi_{kl} + \beta h^{(to)}_{kl}+ \Psi_{lk} + \beta h^{(ot)}_{lk}\Big] \int dt  \big[y_l\hat{f}_k+z_lx_k\big]_{t,\tau}
\end{eqnarray}
 We can now have write  $Z[\bm{\psi},\bm{\Psi}]$ in Eq. \eref{generating_function3a} in terms of an action $S$
$$Z[\bm{\psi}, \bm{\Psi}]=\int D\big[x\hat{x}f\hat{f}h\hat{h}\hat{\mu}wmg yz q\hat{q}Q\hat{Q} K\hat{K}\big]e^{S[x,\hat{x},f,\hat{f}, h,\hat{h}, \hat{\mu}, w, m, g, y, z,q,\hat{q},Q,\hat{Q} ,K,\hat{K}]}$$ 
where
\begin{equation}
S = W_0 +\sum_{k\neq j \in \mathcal{T}} W_{kj}  + \sum_{k \in \mathcal{T}, l \centernot \in \mathcal{T}} \tilde{W}_{kl}  +i\sum_{k \in \mathcal{T}}  \left(\int dt S^{(\psi)}_k(t,\tau) - M_k (\tau) \right)
\end{equation}
\begin{eqnarray} \nonumber W_0 &=\int dt dt'\left\{ q(t,t',\tau)\left[\hat{q}(t,t',\tau) + \frac{i }{2N_t} \,\sum_{l\centernot \in \mathcal{T}} z_l(t,\tau)z_l(t',\tau) \right]\right\} \\\nonumber &+\int dt dt' \left\{Q(t,t',\tau)\left[\hat{Q}(t,t',\tau)  +  \frac{i}{2N_t} \,\sum_{l\centernot \in \mathcal{T}} y_l(t,\tau)y_l(t',\tau)\right]\right\} \\\nonumber &+ \int dt dt' \left\{K(t,t',\tau)\left[ \hat{K}(t,t',\tau)+ \nu\frac{i}{N_t} \,\sum_{l\centernot \in \mathcal{T}}z_l(t,\tau)y_l(t',\tau) \right] \right\} \\ &+  iN_t\int dt \,\hat{\mu}(\tau) m(t,\tau)g(t,\tau) 
\end{eqnarray}
\begin{eqnarray} \nonumber 
M_k &= \hat{\mu}(\tau)  \int dt \big[g(t, \tau) x_k(t,\tau) + m(t,\tau)\hat{f}_k(t,\tau)\big] \\ \nonumber &+ \int dt dt'\big[\hat{q}(t,t',\tau)    x_k(t,\tau)x_k(t',\tau) \big]\\  &+ \int dt dt' \left[\hat{Q}(t,t',\tau)  \hat{f}_k(t,\tau)\hat{f}_k(t',\tau)\ + \hat{K}(t,t',\tau) \hat{f}_k(t',\tau)x_k(t,\tau) \right] 
\end{eqnarray}
\begin{eqnarray} \nonumber
\nonumber W_{kj} &= i\hat{h}_{kj}^{(tt)}\big(h_{kj}^{(tt)}-\tilde{h}_{kj}\big)-  \ln\Big[2{\rm cosh}\, \big(\beta h_{kj}^{(tt)}(\tau)/N_t\big)\Big]  \\ & +  \ln\left[2{\rm cosh}\,\left(\displaystyle \frac{\Psi_{kj}(\tau+1) +\beta h_{kj}^{(tt)}(\tau)- i w_{kj}(\tau)}{N_t} \right)\right] 
\end{eqnarray}
\begin{eqnarray}
\nonumber \tilde{W}_{kl} & = \frac{\big(\Psi_{kl} + \beta h_{kl}^{(to)}+ \Psi_{lk} + \beta h_{lk}^{(ot)}\big)^2}{2N_t} + i\hat{h}_{kl}^{(to)} \big(h_{kl}^{(to)} -\tilde{h}_{kl}\big) \\ \nonumber &  + i\hat{h}_{lk}^{(ot)} \big(h_{lk}^{(ot)} -\tilde{h}_{lk}\big) - \ln\Big[2{\rm cosh}\, \Big(\frac{\beta h_{kl}^{(to)}}{\sqrt{N_t}} \Big)\Big]  - \ln\Big[2{\rm cosh}\, \Big(\frac{\beta h_{lk}^{(ot)}}{\sqrt{N_t}} \Big)\Big]  \\&- i\frac{\Psi_{kl} + \beta h_{kl}^{(to)} + \Psi_{lk} + \beta h_{lk}^{(ot)}}{N_t} \int dt  \big[y_l\hat{f}_k  +  z_l x_k \big]_{t,\tau} 
\end{eqnarray}
Setting $\bm{\Psi}= \mathbf{0}$, saddle-point conditions  wrt $h_{kl}^{(to)}$, $\hat{h}_{kl}^{(to)}$, $h_{lk}^{(ot)}$, $\hat{h}_{lk}^{(ot)}$, $h_{kj}^{(tt)}$ and $\hat{h}_{kj}^{(tt)}$ yield
\numparts 
 \begin{eqnarray*}
 \hat{\lambda}(\tau+1)&\equiv& \sqrt{N_t}  \Big\langle J^{(to)}_{kl}(\tau+1) \Big\rangle := \sqrt{N_t} \left(\lim_{\Psi\rightarrow 0}  \frac{\partial Z}{\partial \Psi_{kl}}\right)  ={\rm tanh}\,\left(\frac{ \beta}{\sqrt{N_t}} h_{kl}^{(to)}(\tau)\right)  \\ \hat{\lambda}(\tau+1)&\equiv&\sqrt{N_t}  \Big\langle J^{(ot)}_{lk}(\tau+1) \Big\rangle := \sqrt{N_t} \left(\lim_{\Psi\rightarrow 0}  \frac{\partial Z}{\partial \Psi_{lk}}\right)  ={\rm tanh}\,\left(\frac{ \beta}{\sqrt{N_t}} h_{lk}^{(ot)}(\tau)\right) \\\hat{\mu}(\tau+1) &\equiv& N_t\Big \langle   J^{(tt)}_{kj}(\tau+1) \Big\rangle  := N_t\left(\lim_{\Psi\rightarrow 0} \frac{\partial Z}{\partial \Psi_{kj}}\right) = {\rm tanh} \left(\frac{ \beta}{N_t} h_{kj}^{(tt)}(\tau)\right)
\end{eqnarray*}
\endnumparts
Next  we   introduce a  measure for the effective dynamics of a single unit $x$ that, for an observable $O = O\big(x, \hat{x}, f, \hat{f}, h_*, \hat{h}_*,   w_*, \hat{\mu}_* \big)$, can be  defined  as
\begin{equation}
\Big\langle O\big(x, \hat{x}, f, \hat{f},h_*, \hat{h}_*,w_*,\hat{\mu}_*\big) \Big\rangle_* = \frac{\displaystyle \int  Dy Dz D\big[x \hat{x} f \hat{f}  \big] O e^{S_*}}{\displaystyle  \int   Dy Dz D\big[x \hat{x} f \hat{f}  \big] e^{S_*} } 
\end{equation}
where $S_*$ denotes the value of the action $S$ evaluated at the saddle point \linebreak $\mathcal{M}_*:=(m_*, g_*,q_*,\hat{q}_*, Q_*, \hat{Q}_*, K_*, \hat{K}_*)$ for $\bm{\psi}= \mathbf{0}$. 

The
saddle-point conditions are:
\begin{equation}
        \left \{ \begin{array}{l} 
        \displaystyle \frac{\partial S}{\partial g(t,\tau)} = 0\,,\quad \rightarrow {m}_*(t, \tau) = \langle x(t; \tau)\rangle_* \\ \\\displaystyle \frac{\partial  S}{\partial q(t,t',\tau)} = 0\,,\quad \rightarrow \hat{q}_*(t,t', \tau) = -\frac{i \alpha}{2}\langle z(t, \tau)z(t, \tau')\rangle_* \\ \\ \displaystyle \frac{\partial  S}{\partial Q(t,t',\tau)} = 0\,,\quad \rightarrow  \hat{Q}_*(t,t', \tau) = -\frac{i\alpha }{2}\langle y(t,\tau)y(t',\tau)\rangle_*\\ \\ \displaystyle  \frac{\partial S}{\partial K(t,t',\tau)} = 0\,,\quad  \rightarrow \hat{K}_*(t,t', \tau) = -i\nu\alpha\langle z(t,\tau)y(t',\tau) \rangle_*
     \end{array} \right.\, 
\end{equation}
\begin{equation}
    \left \{ \begin{array}{l} 
        \displaystyle \frac{\partial S}{\partial m(t,\tau)} = 0\,,\quad \rightarrow {g}_*(t,\tau) = \langle \hat{f}(t,\tau)\rangle_* \\ \\  \displaystyle \frac{\partial S}{\partial \hat{q}(t,t',\tau)} = 0 \,,\quad \rightarrow  q_*(t, t', \tau) = \big\langle  x(t; \tau) x(t',\tau) \big\rangle_* \\ \\\displaystyle \frac{\partial S}{\partial \hat{Q}(t,t',\tau)} = 0  \,,\quad  \rightarrow Q_*(t,t',\tau) = \langle \hat{f}(t,\tau) \hat{f}(t',\tau)\rangle_*  \\ \\ \displaystyle \frac{\partial S}{\partial \hat{K}(t,t',\tau)} = 0  \,,\quad  \rightarrow K_*(t,t',\tau) = \langle x(t, \tau)  \hat{f}(t',\tau) \rangle_*  \end{array} \right.\,,
\end{equation}
 Additionally, from the normalisation $Z[\bm{\psi}= \mathbf{0},\bm{\Psi}= \mathbf{0}] =1$, we have
 $
     \big\langle \hat{f}(t,\tau) \big\rangle_* =0$ and $\big\langle \hat{f}(t,\tau) \hat{f}(t',\tau)\big\rangle_* =0$.
 These imply that $$g_*(t,\tau) = Q_*(t,t',\tau)= 0$$
 
The final step consists of integrating out $y$ and $z$ to find $\hat{q}_*,   \hat{Q}_*,  \hat{K}_* $ as follows
\begin{eqnarray}
   \nonumber \hat{q}_*(t,t',\tau) &= -\frac{i \alpha }{2 }\left\langle\left(\sum_{\ell'\in \mathcal{T}} J_{\ell'}^{(to)}(\tau)\hat{f}_{\ell'}(t,\tau) \right)\left(\sum_{\ell \in \mathcal{T}}J_{\ell}^{(to)}(\tau)\hat{f}_{\ell}(t',\tau)  \right) \right\rangle_*\\
\nonumber &= -\frac{i\alpha}{ 2} \sum_{\ell \in\mathcal{T}} \left\langle \big[J_{\ell}^{(to)}(\tau)\big]^2  \hat{f}_{\ell}(t,\tau)  \hat{f}_{\ell}(t',\tau)\right \rangle_*  \\ &= 0 \\
\nonumber
    \hat{Q}_*(t,t',\tau) &= -\frac{i \alpha}{2 }\left\langle\left(\sum_{\ell'\in \mathcal{T}} J_{\ell'}^{(ot)}( \tau)x_{\ell'}(t,\tau)  \right)\left(\sum_{\ell\in \mathcal{T}} J_{\ell}^{(ot)}( \tau) x_{\ell}(t',\tau)  \right) \right\rangle_* \\ \nonumber &= -\frac{i\alpha}{ 2}\sum_{\ell} \left\langle \big[J_{\ell}^{(ot)}( \tau)\big]^2   x_{\ell}(t,\tau)  x_{\ell}(t',\tau)\right \rangle_*  \\ &=- \frac{i\alpha  }{2 } C(t,t',\tau)
\\ \nonumber
    \hat{K}_*(t,t',\tau) &= - i \nu\alpha\left\langle\left(\sum_{\ell'\in \mathcal{T}} J_{\ell'}^{(to)}( \tau)\hat{f}_{\ell'}(t, \tau)  \right)\left(\sum_{\ell\in \mathcal{T}} J_{\ell}^{(ot)}( \tau)x_{\ell}(t', \tau)  \right) \right\rangle_* \\ \nonumber &= -i\nu\alpha  \sum_{\ell} \left\langle J_{\ell}^{(to)}( \tau)J_{\ell}^{(ot)}( \tau)\right \rangle_* K(t',t, \tau)  \\ &= - i\nu\alpha\,\Big(iG(t',t, \tau)\Big)
\end{eqnarray}
where $ K(t',t, \tau) =\left\langle \hat{f}_{\ell}(t,  \tau)  x_{\ell}(t',  \tau)\right \rangle_*$ and $G(t',t,  \tau):=\partial \langle x(t',  \tau)\rangle_*/\partial \theta(t, \tau)$.

The effective single-unit dynamics is generated by the following path probability:
\begin{eqnarray} \mathcal{P}\big(\{x\}, \{f\}\big) & = \int D\hat{x}D\hat{f} \prod_{\tau =0}^{T_{\rm max}}\exp\left\{ i\int dt \Big[ \hat{x} \big(\partial_t + 1 \big)x +i\frac{\sigma^2}{2} \hat{x}^2 - \hat{x}\,F(f )  \Big] \right\} \\  \nonumber &\times  \exp\left\{ i\int dt\hat{f}\big(f-  m_*\hat{\mu}(\tau)  - \theta \big)\right\}\\\nonumber &\times \exp\left\{- \alpha  \int dt dt' \Big[  \frac{C(t,t', \tau)}{2} \hat{f}(t,\tau)\hat{f}(t',\tau) + i\nu G(t',t,\tau)x(t; \tau)\hat{f}(t',\tau)\Big]\Big\}\right\}
\end{eqnarray}
We can further perform integration over the auxiliary fields $\{\hat{f}\}$, using  a Gaussian noise $\eta(t,\tau)$  with correlator $\langle \eta(t,\tau)\eta(t', \tau)\rangle = \alpha C(t,t', \tau)$, to rewrite this path probability as
\begin{eqnarray}
\nonumber \mathcal{P}\big(\{x\}, \{f\}\big)   &=  \int D\hat{f} \left. \exp\Big\{-  \frac{\alpha }{2} \int dt dt'  \,  C(t,t',\tau) \hat{f}(t,\tau)\hat{f}(t',\tau) +i\int dt \hat{f}(t,\tau)\eta(t,\tau)\Big\}\right] \\ \nonumber  & \times \int D \xi P(\xi) D\eta \prod_{\tau =0}^{T_{\rm max}} \left\{\delta\Big[  \big(\partial_t + 1 \big)x  -\,F(f ) - \xi\Big]  \right.\\ &\times \left.\delta\left(f-    m_*\hat{\mu}(\tau) -  \alpha\nu\int dt' G(t,t',\tau)x(t',\tau)  - \eta- \theta\right)  \right\} \end{eqnarray}
Finally,
\begin{eqnarray}
\nonumber 
\mathcal{P}\big(\{x\}, \{f\}\big)  &= \int D\xi P(\xi)D\eta P(\eta) \prod_{\tau =0}^{T_{\rm max}}   \delta\big[  \big(\partial_t + 1 \big)x  -\,F(f ) - \xi\big] \\ &\times \delta\Big(f-    m_*\hat{\mu}(\tau) -  \alpha\nu\int dt' G(t,t',\tau)x(t')  - \eta- \theta  \Big)
\end{eqnarray}
 The above equation results  in the equivalent SDE form of the $x$'s dynamics in \Eref{closed1}. We note that the single-unit effective measure now can also be defined as
 \begin{eqnarray}
     \langle O(\{x\}) \rangle_* &= \int Dx\left[\int  D\xi P(\xi) D\eta P(\eta)   \mathcal{P}\big(\{x\}\big|\{\eta\}, \{\xi\}\big)\right]  O(\{x\})\\ &= \int  D\xi P(\xi) D\eta P(\eta) O\big(\{x\}\big|\{\eta\}, \{\xi\}\big)
\end{eqnarray}
 where the conditional average $O\big(\{x\}\big|\{\eta\}, \{\xi\}\big)$  is taken wrt the conditional probability
  \begin{eqnarray*}
    \mathcal{P}\big(\{x\}
    \big|\{\eta\}, \{\xi\}\big):= \prod_{\tau=0}^{T_{\rm max}} \prod_t\Big\langle \delta\big(\partial_t x(t, \tau) + x(t, \tau) - F(f(t,\tau))  - \xi(t,\tau) \big) \Big\rangle_*\Big|_{\displaystyle f= \kappa } \\ \kappa  = m_*(t, \tau)\hat{\mu}(\tau) + \alpha \nu \int dt' G(t,t',\tau)x(t',\tau) +\eta(t, \tau) +\theta(t, \tau)
\end{eqnarray*}
 \subsection*{D.  Derivation of  \Eref{largesteigenvalue} for the outlier in the spectrum of the target-gene effective dynamics' Jacobian around the zero fixed point}
 In this section we consider only the coupling matrix $\mathbf{J}$ that is achieved at the infinite generation $\tau \rightarrow \infty$. So all the $\tau$-dependence can be dropped out in considering the $x$'s dynamics as well as in the notation of $\mathbf{J}$. 
 We have
 $ \hat{\mu}(\tau) := \big\langle J_{ij}^{(tt)}(\tau)\big\rangle \rightarrow  \mu/N_t$ and   $ \hat{\lambda}(\tau) := \big\langle J_{ik}^{(to)}\big\rangle   =\big \langle J_{ki}^{(ot)}\big\rangle \rightarrow  \lambda/\sqrt{N_t}$ as $\tau \rightarrow \infty$. Denoting by $\underline{\mathbf{J}}^{(tt)}$ the set of effective interactions $J_{ij}^{(e)}$ as defined in Eq. \eref{signal_to_noise}, we can consider  $J_{ij}^{(e)}$ as as Gaussian random variable whose mean is $\mu/N_t$ and variance is $\lambda^4/N_t$ with a covariance between $J_{ij}^{(e)}$ and $J_{ji}^{(e)}$ quantified by a symmetry parameter $\Gamma= \nu^2 \in [0,1]$:  
\begin{equation}
J_{ij}^{(e)}=  \mathcal{N}\left(\frac{\mu}{N_t}, \frac{\lambda^4}{N_t} \right) \,,\qquad {\rm and} \,\big[J_{ij}^{(e)},  J_{ji} ^{(e)}\big]_{\underline{\mathbf{J}}} = \frac{\Gamma \lambda^4}{N_t} 
 \end{equation}
where we have used $\big[\cdot\big]_{\underline{\mathbf{J}}}$ to denote the average taken with respect to the ensemble of random realisations of $\underline{\mathbf{J}}^{(tt)}$  and $\big[a,b\big]_{\underline{\mathbf{J}}}$ to denote the covariance of $a$ and $b$.  In order to compute the outlier of spectrum of  $\underline{\mathbf{J}}^{(tt)}$,  we follow the approach described in \cite{Baron2023}, which shows that the outlier of $\underline{\mathbf{J}}^{(tt)}$, for $\mu \neq 0$,  needs to satisfy \cite{BENAYCHGEORGES, Orourke}
  \begin{equation}
  R(1+ \omega_{\rm outlier}) = \frac{1}{\mu}\,,\qquad R(\omega) := \frac{1}{N_t}\Big[\sum_{i,j} R_{ij}(\omega) \Big]_{\underline{\mathbf{J}}}
 \end{equation}
 where $R_{ij}$ are the entries of the resolvent  $\mathbf{R}(\omega) = \big(\omega\mathbf{I}_{N_t}- \mathbf{z}\big)^{-1}$, $z_{ij} = J_{ij}^{(e)} - \mu/N_t$ and $\mathbf{I}_{N_t}$ is the $N_t$-by-$N_t$ identity matrix.
 Using the Neumann series for $\mathbf{R}$, we have
 \begin{equation}
     R(\omega) = \frac{1}{N_t}\Big[\sum_{i,j} \big(\omega \delta_{ij} - z_{ij}\big)^{-1} \Big]_{\underline{\mathbf{J}}}= \frac{1}{N_t}\left[\sum_{i,j} \left(\frac{\delta_{ij}}{\omega} + \frac{z_{ij}}{\omega^2} + \sum_k \frac{z_{ik}z_{kj}}{\omega^3}  + \cdots \right) \right]_{\underline{\mathbf{J}}}
     \label{resolvent1}
 \end{equation}
 We introduce the moment generating functional $\underline{Z}[\bm{\phi}]$ for the dynamics of Eq. \eref{target_dynamics}:    $$\underline{Z}[\bm{\phi}] = \int D\big[x\hat{x} f\hat{f}\big]\exp \Big\{i \sum_{k\in \mathcal{T}}\int dt \tilde{S}_k^{(0)}(t)- i \sum_{k,j\in \mathcal{T}} \int dt \big[ \phi_{kj}(t) z_{kj}(t) + J^{(e)}_{kj} x_j(t) \hat{f}_k(t)\big]\Big\}$$
where $\tilde{S}_k^{(0)}(t) = \lim_{\tau \rightarrow \infty} S_k^{(0)}(t,\tau)$ with $S_k^{(0)}(t,\tau)$ given by Eq. \eref{S_k}. When taking the average  $\Big[\underline{Z}[\bm{\phi}]\Big]_{\underline{\mathbf{J}}}$, one needs to compute  the average of the second exponent
\begin{eqnarray*} H[\bm{\phi}]&= \left[\exp\left\{ - i \sum_{k,j\in \mathcal{T}} \int dt \big[ \phi_{kj}(t) z_{kj}(t) + J^{(e)}_{kj} x_j(t) \hat{f}_k(t)\big]\right\} \right]_{\underline{\mathbf{J}}} \\ &\propto \exp\left\{ -\frac{\lambda^4}{2N_t}\sum_{i,j} \left[\int dt \big(\phi_{ij} + x_j\hat{f}_i\big)\right]^2 \right\} \\ & \times \exp\left\{-\frac{\Gamma \lambda^4}{2N_t}\sum_{i,j}  \int dt dt' \Big[\phi_{ij}(t) + x_j(t)\hat{f}_i(t)\Big]\cdot\Big[\phi_{ji}(t') + x_i(t')\hat{f}_j(t')\Big] \right\} 
\end{eqnarray*}
 Denoting
 \begin{equation}
     V_{ij} :=\frac{\delta (\ln H)}{\delta \phi_{ij}(t)} = -\frac{\lambda^4}{N_t}\left\{ \int dt' \Big[\phi_{ij}(t') + x_j\hat{f}_i(t')\Big]  + \Gamma\int dt'  \Big[\phi_{ji}(t') + x_i(t')\hat{f}_j(t')\Big] \right\}
 \end{equation}
 we can express the moments of $z_{ij}$ via the  derivatives of $H[\bm{\phi}]$ averaged over the dynamical realisations of $\mathbf{x}$ (this kind of averages is denoted by $\langle \cdot \rangle$) as, for example,
$$     \big[z_{ij}\big]_{\underline{\mathbf{J}}} = i \frac{\delta Z}{\delta \phi_{ij}}\Big|_{\bm{\phi}=0} = i \left\langle  \frac{\delta \ln H}{\delta \phi_{ij}} \right \rangle\Big|_{\bm{\phi}=0}= i \big\langle V_{ij} \big\rangle\big|_{\bm{\phi}=0}
$$
This allows one to rewrite Eq. \eref{resolvent1} as
\numparts 
  \begin{eqnarray*}
           R(\omega) & = \frac{1}{N_t}\sum_{i,j} \left[\frac{\delta_{ij}}{\omega} + \frac{i}{\omega^2}\,  \left\langle  V_{ij} \right \rangle- \frac{1}{\omega^3} \sum_k \left\langle V_{ik}V_{kj} + \frac{\delta V_{ik}}{\delta \phi_{kj}}   \right \rangle \right.\\ &-\left.\frac{i}{\omega^4} \sum_{k,l} \left\langle V_{ik}V_{kl}V_{lj} + V_{ik}\frac{\delta V_{kl}}{\delta \phi_{lj}}  + \frac{\delta V_{ik}}{\delta \phi_{kl}} V_{lj}  \right \rangle + \cdots \right]_{\bm{\phi}=0}
  \end{eqnarray*}
\endnumparts 
We remark the two properties of $V_{ik}$ (terms that are of order $O(N^{-1})$ are neglected), namely
\begin{equation}
    \left \{ \begin{array}{l} \displaystyle \frac{1}{N_t}\sum_{i,j,k} \left\langle \frac{\delta V_{ik}}{\delta \phi_{kj}} \right\rangle_{\bm{\phi}=0}= -\frac{\Gamma \lambda^4}{N_t^2} \sum_{i,j,k}  \delta_{ij} = -\Gamma \lambda^4 \\ \\ \displaystyle \frac{1}{N_t}\sum_{i,j,k,l} \left\langle V_{ik}\frac{\delta V_{kl}}{\delta \phi_{lj}}
 \right\rangle_{\bm{\phi}=0} =-\frac{\Gamma \lambda^4}{N_t} \sum_{i,j} \big\langle  V_{ij} \big \rangle\big|_{\bm{\phi}=0}\end{array} \right.\,
\end{equation}
After some calculations following the procedure detailed in \cite{Baron2023}, we find
\begin{equation}
    R(\omega) = u(\omega) + v(\omega) 
\end{equation}
 where 
    $$u(\omega) = \frac{1}{\omega} - \frac{1}{N\omega^3}\sum_{i,j,k} \left\langle\frac{\delta V_{ik}}{\delta \phi_{kj}}\right\rangle_{\bm{\phi}=0} + \frac{1}{N\omega^5}\sum_{i,j,k,l,m} \left\langle  \frac{\delta V_{ik}}{\delta \phi_{kl}}\frac{\delta V_{lm}}{\delta \phi_{mj}}  + \frac{\delta V_{ik}}{\delta \phi_{mj}} \frac{\delta V_{lm}}{\delta \phi_{kl}}  \right\rangle_{\bm{\phi}=0} +\cdots$$
 $$  v(\omega) = \frac{iu^2}{N}\sum_{i,j} \big\langle  V_{ij} \big \rangle\big|_{\bm{\phi}=0} - \frac{u^3}{N}\sum_{i,j,k} \big\langle  V_{ik}V_{kj} \big \rangle\big|_{\bm{\phi}=0} - \frac{iu^4}{N}\sum_{i,j,k,l} \big\langle  V_{ik}V_{kl}V_{lj} \big \rangle\big|_{\bm{\phi}=0} +\cdots$$
The function $u(\omega)$ can be obtained as a solution to the equation \cite{Baron2023}
\begin{equation}
    u(\omega) = \frac{\omega -\sqrt{\omega^2 -4\Gamma \lambda^4 }}{2\Gamma \lambda^4}
\end{equation}
while for $v(\omega)$ we have
\begin{equation}
    v(\omega) = \frac{u^2 \Big[(1+\Gamma)m \hat{f}_*+ \Gamma u\big(Q_* m^2 +  q_*\hat{f}_*^2 - 2 \chi m \hat{f}_* \big)\Big]}{1-(1+\Gamma)\chi u + \Gamma u^2 (\chi^2 - q_*Q_*)}
\end{equation}
Since $\hat{f}_* := \sum_{k} \hat{f}_k/N =0$ and $ Q_* =0$, $v(\omega) =0$. Therefore,
\begin{equation}
    R(1+\omega_{\rm  outlier}) = \frac{\omega_{\rm  outlier}+ 1 -\sqrt{\big(\omega_{\rm  outlier}+1\big)^2 -4\Gamma\lambda^4 }}{2\Gamma \lambda^4} =\frac{1}{\mu}
\end{equation}
The solution to this equation, for $\Gamma=1$ (i.e. fully symmetric interactions), reads
\begin{equation}
    \omega_{\rm  outlier} = \frac{ \lambda^4+ \mu^2}{\mu} - 1
       \label{outlier}
\end{equation}
This equation shows when $\mu = 1$, $\omega_{\rm outlier} = \lambda^4$. Since the Jacobian for the dynamics of target genes around the zero fixed point $\mathbf{x}=\mathbf{0}$ is $\mathcal{J}_{ij}= - \delta_{ij} + \underline{J}_{ij}^{(tt)} $. The largest eigenvalue of $\mathcal{J}$ is hence $\lambda^4-1$. In the case of fully symmetric interactions,  the zero fixed point thus becomes marginally stable  in the robust phase where  both $\mu,\lambda \rightarrow 1$. 
As a final remark, while it is possible to  extend the above computation of the outlier to the more general case of non-zero fixed point, we leave it for future work. 

\subsection*{E. Linear stability of the  fixed point  in \Eref{fixed_point_of_closed1}  in the absence of noise}
 
 Here we follow the approach of \cite{Opper1992}, where  we perturb the fixed point $x_*$ by some small Gaussian white noise  and check how the system responds to it. To this end, we linearize the ADMFT second  Eq.  \eref{closed1} around $x_*$ and perform a Fourier analysis as follows.
For $\sigma =0$, with  $z=\mathcal{N}(0,1)$, Eq.  \eref{fixed_point_of_closed1} then becomes
 \begin{equation}
 x_*(z) =  f_0(x_*) :={\rm tanh}\Big(\mu m_\infty +  J_0\sqrt{q}z +  \chi \nu\alpha  x_*\Big) 
\label{fixed_point_no_noise}
 \end{equation}
where $m_\infty=\langle x_*\rangle_z$ and $q=\langle x_*^2\rangle_z$.
For $\mu=0$, $x_* =0$ is always a fixed point.
Perturbations around any fixed point $x_*$ can be described by $x(t) = x_*+\varepsilon x_1(t)$ and  $\eta(t) = J_0\sqrt{q}z + \varepsilon z_1(t)$. We have $\langle z_1(t)z_1(t') \rangle = \alpha \langle x_1(t)x_1(t') \rangle $ from the self-consitency relation Eq. \eref{self_consistency_noise}. Adding a term $\varepsilon \phi(t)$ into the  effective process Eq. \eref{closed1} for $\sigma=0$, where  $\phi(t)$ is a white noise   of unit variance, the  Fourier components  $\tilde{x}_1(\omega)$ of $x_1(t)$ satisfy
\begin{equation}  \tilde{x}_1 = \frac{\big[1-f_0^2(x_*)\big]\tilde{\phi}(\omega)}{i\omega +1 -\nu \alpha\big[1-f_0^2(x_*)\big]\tilde{G}(\omega)} \label{Fourier1}\,, \end{equation}
Solving this equation for the square modulus of $\tilde{x}_1(\omega)$, i.e. $\tilde{R}_0(\omega) = \big|\tilde{x}_1(\omega) \big|^2$ gives us
\begin{equation}
 \tilde{R}_0^{-1}(\omega) = \left|\frac{(i\omega +1) -\nu \alpha\tilde{G}(\omega) \big[1-f_0^2(x_*)\big] }{1-f_0^2(x_*)}\right|^2 -  \alpha
\end{equation}
The stability of the steady-state solution in Eq. \eref{fixed_point_no_noise} is only determined by $\omega=0$. In this case, $\tilde{G}(\omega=0) = \chi$ and the above equation simplifies to \begin{equation}
\big[\tilde{R}_0(\omega=0) \big]^{-1}= \frac{\Big[1- \nu \alpha \chi(1-f_0^2(x_*) \big) \Big]^2}{\big[1-f_0^2(x_*)\big]^2} -  \alpha
\end{equation}
 If $\tilde{R}_0(\omega=0) $ diverges, perturbations do not decay to zero, signalling an instability of the fixed point. This means the fixed point $x_*$ loses its stability at $\Sigma =0$, while remains stable at  $\Sigma >0$, for
\begin{equation}
    \Sigma(x_*) = \Big[1- \nu \alpha\chi(1-f_0^2(x_*) \big) \Big]^2 - \alpha\big[1-f_0^2(x_*)\big]^2 
    \label{stability}
    \end{equation}
For the paramagnetic fixed point with $x_*=q=0$, so that $f_0(x_*)=0$, this condition simplifies to 
\begin{equation}
    \Sigma(x_* =0) =\big[1-  \nu \alpha\chi \big]^2 -  \alpha   \label{stability_para}
\end{equation}
For $\nu=0$ this reduces to the well-known result $\alpha_c=1$  for static random networks \cite{Sompolinsky}, while for $\nu\neq 0$ we can 
proceed with the definition of $\chi$
\begin{equation}
    \chi = \frac{1}{J_0 \sqrt{q}}\left\langle \frac{\partial x_*(z)}{\partial z}\right \rangle_z= \displaystyle \left\langle \frac{1-f_0^2(x_*)}{1-\nu \alpha\chi \big(1-f_0^2(x_*)\big)} \right\rangle_z
\end{equation}
that, around the critical line where $x_*\simeq 0$ and hence $m_\infty$ and $q$ are small, satisfies the following quadratic equation:
\begin{equation}
    \chi(1-\chi \nu \alpha) = 1
\end{equation}
The \emph{physical} solution to this equation reads
\begin{equation}
    \chi_* = \frac{1-\sqrt{1-4\nu \alpha}}{2\nu \alpha}
\end{equation}
Substituting this $\chi_*$ into Eq. \eref{stability_para}, we arrive at
\begin{equation}
   \Sigma(x_* =0) = \frac{1}{4}\Big[1+ \sqrt{1-4\nu \alpha}\Big]^2 -\alpha
\end{equation}
It is easy to check that this reduces to $\alpha_c(1+\nu)=1$ and hence $\alpha_c =0.5$ for $\nu=1$.
\section*{References}

\bibliographystyle{unsrt}
\bibliography{Doublereplica}
\end{document}